\documentclass[english,osajnl,twocolumn,showpacs,superscriptaddress,10pt]{revtex4-1}
\usepackage[T1]{fontenc}
\usepackage[latin9]{inputenc}
\usepackage{color}
\usepackage{amsmath}
\usepackage{amssymb}
\usepackage{graphicx}
\usepackage{esint}
\usepackage{babel}
\begin{document}

\title{Decoherence of Einstein-Podolsky-Rosen steering}

\author{L. Rosales-Z\'arate, R. Y. Teh, S. Kiesewetter, A. Brolis, K. Ng
and M. D. Reid }

\affiliation{Centre for Quantum and Optical Science, Swinburne University of Technology,
Melbourne, 3122 Australia}
\begin{abstract}
We consider two systems $A$ and $B$ that share Einstein-Podolsky-Rosen
(EPR) steering correlations and study how these correlations will
decay, when each of the systems are independently coupled to a reservoir.
EPR steering is a directional form of entanglement, and the measure
of steering can change depending on whether the system $A$ is steered
by $B$, or vice versa. First, we examine the decay of the steering
correlations of the two-mode squeezed state. We find that if the system
$B$ is coupled to a reservoir, then the decoherence of the steering
of $A$ by $B$ is particularly marked, to the extent that there is
a sudden death of steering after a finite time. We find a different
directional effect, if the reservoirs are thermally excited. Second,
we study the decoherence of the steering of a Schrodinger cat state,
modelled as the entangled state of a spin and harmonic oscillator,
when the macroscopic system (the cat) is coupled to a reservoir.

OCIS: 270.0270; 270.6570; 270.5568
\end{abstract}
\maketitle

\section{Introduction}

The sort of nonlocality we call ``Einstein-Podolsky-Rosen-steering''
\cite{Schrodinger,key-2,hw-steering-1,hw2-steering-1,EPRsteering-1,eprsteerphoton-1}
originated in 1935 with the Einstein-Podolsky-Rosen (EPR) paradox
\cite{epr}. The EPR paradox is the argument that was put forward
by Einstein, Podolsky and Rosen for the incompleteness of quantum
mechanics. The argument was based on premises (sometimes called \emph{Local
Realism} or in Einstein's language, no ``spooky action-at-a-distance'')
that were not restricted to classical mechanics, but were thought
essential to any physical theory \cite{bell book}. The EPR argument
reveals the inconsistency between Local Realism and the completeness
of quantum mechanics. It does not in itself rule out all completions
of quantum mechanics, that are compatible with local realism. Nowadays,
after the work of Bell, we realise this ruling out of all local realistic
theories can be done in some special experimental situations \cite{Bell,key-4,bellloophole,key-3}.
Any realisation of the EPR paradox \cite{wuepr,rmp-1,ou}, as a special
case of EPR steering, is nonetheless important in giving a concrete
intermediate result: the fact that quantum mechanics without completion
cannot be consistent with local realism. A great advantage of EPR-steering
over Bell-type tests is that they are more accessible to mesoscopic
or macroscopic systems. EPR steering tests have been quantitatively
studied or proposed for optical down conversion \cite{ou,mdrdrumm,rmp-1,mdrepr-1},
optical systems in nonlinear regimes near or at critical points \cite{karl},
atomic gas ensembles at room temperature \cite{polzik room temp,thermal atomic ensembles,key-5},
Bose-Einstein condensates (BEC) \cite{eprbecobert,karenbec,beceprbargill,key-6,key-7,key-8,ferriseprbec,prlobertbec,key-9,key-10},
and opto-mechanics \cite{simon,opto,paoloepr,thermal he andR}. This
may lead to experimental tests of Schrodinger cat-type states.

\begin{figure}
\begin{centering}

\par\end{centering}

\begin{centering}
\includegraphics[width=0.75\columnwidth]{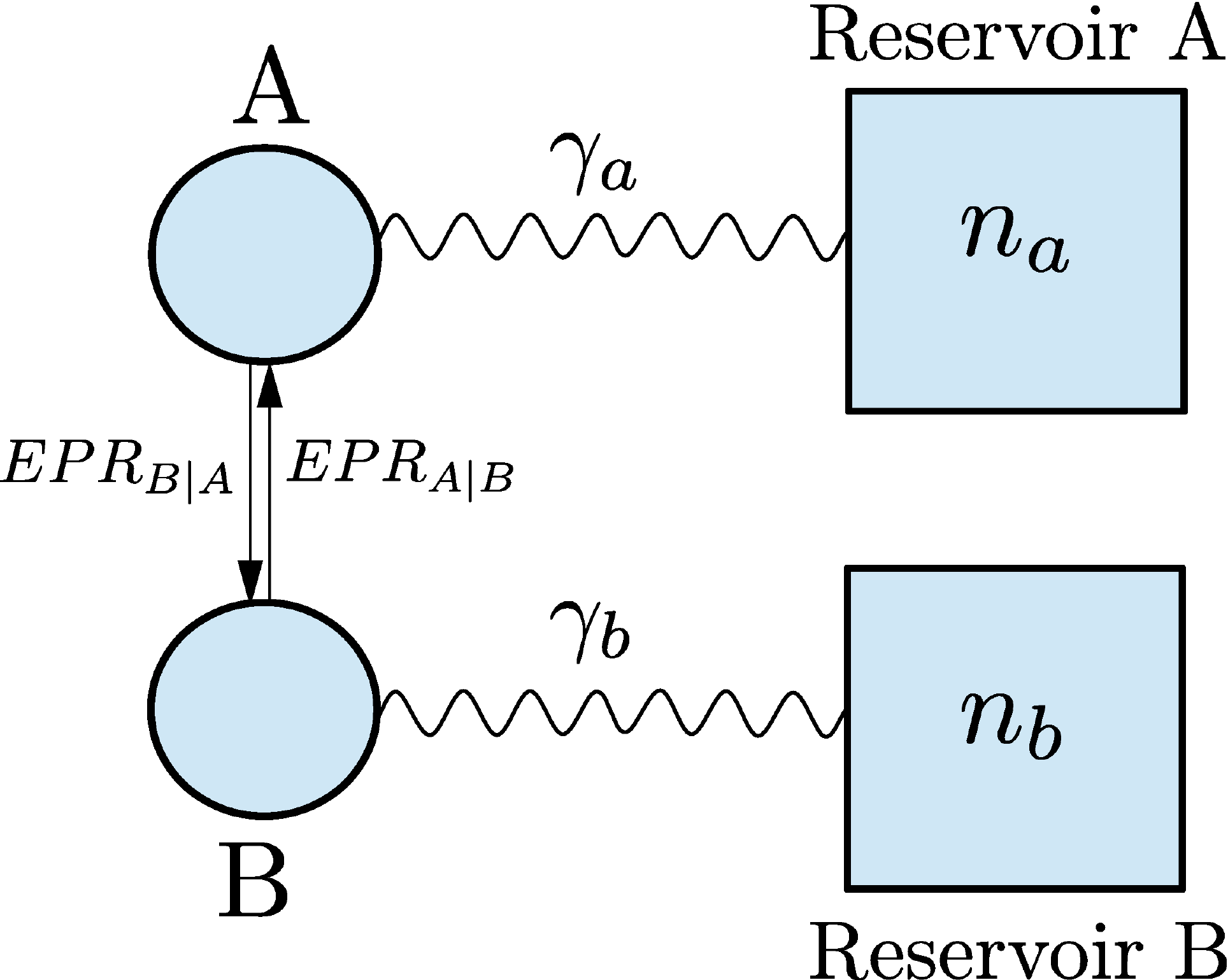}
\par\end{centering}

\protect\caption{\emph{\label{fig1}}Our question is this: We consider two well-separated
systems, $A$ and $B$, that possess the type of quantum correlation
we call EPR-steering. Each system is independently coupled to a thermal
reservoir, at a time $t=0$. This coupling induces a decay of the
EPR-steering. We ask how does the decay of the steering depend on:
the coupling $\gamma_{a}$, $\gamma_{b}$ to each reservoir; the thermal
excitation $n_{a}$, $n_{b}$ of each reservoir; and the measure that
gives the strength of the EPR-steering. We will consider two types
of EPR systems: The first is a two-mode squeezed state. The second
is a mesoscopic/ macroscopic entangled ``Schrodinger cat'' state.\textcolor{blue}{}}
\end{figure}

The EPR paradox manifests as a strong correlation between the positions
and momenta of two spatially separated particles (or some equivalent
correlations). These correlations, once one assumes Local Realism,
become inconsistent with the uncertainty principle. Given the strangeness
of the EPR correlations, a likely hypothesis would be that they do
not exist in real physical systems. Indeed, the sort of correlations
needed for EPR are not easily realisable in experiment. To date, only
a few experiments can justify the claim of loophole-free EPR correlations
\cite{witt NJP zeilloopholefree-1}, or steering without detection
efficiency loopholes \cite{loopholesteer,smith UQ steer -1,ou,rmp-1}.
This paper examines why this is so: There are two possibilities. 

(1) It could be that quantum mechanics needs modification to the extent
that EPR correlations cannot be predicted. If we accept irrefutable
evidence now exists for EPR correlations, then this probability would
appear falsified. However, we comment that the experimental realisations
for EPR steering have been for optical systems, not yet for atoms
or mesoscopic devices.

(2) Or else: a popular belief is that quantum mechanics is correct
as it is, and predicts exactly why, to a great accuracy, the EPR correlations
are extraordinarily difficult to measure. In that case, these predictions
need to be evaluated and verified by experiment.

This last point is what we examine in this paper: We study in detail
why, according to quantum predictions, the EPR correlations are not
easily realisable. As with the well known ``Schrodinger cat'' \cite{schcat,key-14,catstrappedions,key-11,singlemodecat,cats,cats3,yurke stoler,key-12,key-13},
we expect this is so, because of the effect that occurs when a quantum
system is coupled to its environment, modelled as a large system -
a reservoir \cite{legdecoh}. This effect is called ``decoherence''. 

Here, we examine some different sorts of decoherence that occur for
EPR steering. As with the decay of entanglement, there are many cases
one can study. It is worth noting that EPR steering was realised using
high efficiency detection for optical amplitudes \cite{ou,rmp-1},
and has also been unambiguously detected in photonic scenarios with
exceptional losses ($\sim87\%$) \cite{loopholesteer}. \textcolor{blue}{}Tests
have also begun for its genuine multipartite form, distributed among
different locations \cite{seijigen,key-17,key-15,key-16}. 

In this paper, we restrict our investigation to the following cases:
We study position-momentum measurements or their optical equivalent,
quadrature phase amplitude measurements. First, we analyze the steering
of a two-mode squeezed state, to understand the decoherence effects
of coupling to a thermal reservoir. We examine two properties of the
decoherence: the effect of simple losses (damping); and the effect
of thermal noise. Second, we consider a special case of EPR-steering,
that relates to a macroscopic/mesoscopic superposition state - a ``Schrodinger
cat'' state. In fact, the EPR-steering paradox can be used to signify
the Schrodinger cat superposition \cite{erci cav reid pra }. We examine
the effect of decoherence as caused by interaction with the environment,
on this signature. 

An interesting new feature evident for EPR-steering is that the nonlocality
can manifest \emph{asymmetrically} with respect to the observers \cite{steer thmurray,brunner,onewaysteer,eprsteer-1,asymmschnee,monog}.
In our results, we focus on the asymmetry of the decoherence effects
on the EPR-steering. This is valuable to understand, not only from
the fundamental perspective of testing quantum mechanics, but from
the point of view of the potential applications of EPR-steering, which
include cryptography \cite{epr crypto,norbet,key-18,key-19,onesided,grangiertele}
and no-cloning teleportation \cite{grangiertele}. One-sided device-independent
cryptography has been proposed using EPR steering inequalities \cite{onesided}.
In recent papers, it was shown how the asymmetric effects of loss
on EPR-steering could affect the optimal location of teleportation
stations \cite{tele}, or which entanglement criterion should be used
to faithfully verify entanglement in cases of untrusted devices \cite{onesided,onesidedbog,grangiertele,norbet,tele}.

\section{Decoherence of steering for the two-mode squeezed state}

Consider two quantum harmonic oscillators, with boson operators $a$
and $b$, and depicted in the Figure 1 as systems $A$ and $B$. Such
oscillators can be coupled, so that EPR-steering correlations are
induced between the two oscillators.

One of the nicest examples of such EPR steering has been realised
experimentally in optics \cite{rmp-1} as the two-mode squeezed state
\cite{tmss-1}. Here, each mode of the light field is modeled as a
quantum harmonic oscillator. An example of a coupling between the
oscillators that generates this type of state is called parametric
down conversion, and can be described by the interaction Hamiltonian
in an appropriate rotating frame \cite{mdrepr-1}
\begin{equation}
H=i\kappa E(ab-a^{\dagger}b^{\dagger})\label{eq:Hmatwo-modesq}
\end{equation}
Similar two-mode squeezed states can be generated by impinging two
single-mode squeezed states into the two different input ports of
an optical beam splitter \cite{tele-1}. 

The two-mode squeezed state system is fundamentally interesting, as
it can be quite accurately realised in optics, but also it gives a
reasonable approximation to the effects we expect to see in many other
physical realisations of steering, such as in opto-mechanics \cite{opto,thermal he andR,simon},
atomic ensembles \cite{polzik room temp,thermal atomic ensembles,key-5},
and BEC \cite{eprbecobert,beceprbargill,key-6,key-7,key-8,ferriseprbec,prlobertbec,key-9,key-10}.
The study of the effect of the thermal environment on the EPR steering
for a two-mode squeezed state will therefore give us insight into
a broad set of physical scenarios. The two-mode quantum correlation
effects were noticed originally in the context of photon number correlations
between the two beams, which gives rise to noise reduction \cite{heidmann-1,key-20}.

\subsection{The two-mode squeezing Hamiltonian}

\textcolor{red}{}We begin by reviewing the solution for the EPR correlations
given by the Hamiltonian of Eq. (\ref{eq:Hmatwo-modesq}). The EPR
solutions were originally derived in \cite{mdrepr-1}. We define the
quadrature phase amplitudes $X_{A}$, $P_{A}$, $X_{B}$ and $P_{B}$
for each mode: $X_{A}=a+a^{\dagger},P_{A}=(a-a^{\dagger})/i$ and
$X_{B}=b+b^{\dagger},P_{B}=(b-b^{\dagger})/i$. This choice of scaling
for the amplitudes will imply the Heisenberg uncertainty relations
$\Delta X_{A}\Delta P_{A}\geq1$ and $\Delta X_{B}\Delta P_{B}\geq1$.
We note that depending on the physical system modelled by the Hamiltonian,
the amplitudes can also correspond to scaled position and momentum
observables. We now ask what correlations are generated between the
quadrature amplitudes, after an interaction time $\tau$ between the
two modes, at sites denoted by $A$ and $B$.

On examining (\ref{eq:Hmatwo-modesq}), the resulting coupled equations
for $a$ and $b$ can be readily solved, to give\textcolor{black}{
\begin{eqnarray}
a(\tau) & = & \eta a(0)-\sqrt{(\eta^{2}-1)}b^{\dagger}(0)\nonumber \\
b(\tau) & = & \eta b(0)-\sqrt{(\eta^{2}-1)}a^{\dagger}(0)\label{eq:solnarray1}
\end{eqnarray}
}where $\eta=\cosh r$.  We define the squeezing parameter as $r=\frac{Ek}{\hbar}\tau$.
The quadrature phase amplitudes are given by
\begin{eqnarray}
X_{A}\left(\tau\right) & = & \cosh rX_{A}\left(0\right)-\sinh rX_{B}\left(0\right)\nonumber \\
P_{A}\left(\tau\right) & = & \cosh rP_{A}\left(0\right)+\sinh rP_{B}\left(0\right)\nonumber \\
X_{B}\left(\tau\right) & = & \cosh rX_{B}\left(0\right)-\sinh rX_{A}\left(0\right)\nonumber \\
P_{B}\left(\tau\right) & = & \cosh rP_{B}\left(0\right)+\sinh rP_{A}\left(0\right)\label{eq:arraysolns}
\end{eqnarray}
This enables us to calculate correlations such as $\langle X_{A}^{2}\rangle$,
$\langle X_{A}X_{B}\rangle$ after a time $\tau$, assuming initial
uncorrelated vacuum states for all modes. We can then evaluate
the variance of $X_{A}-g_{x}X_{B}$, which will be denoted $\Delta^{2}\left(X_{A}-g_{x}X_{B}\right)$
and which equals $\left\langle \left(X_{A}-g_{x}X_{B}\right)^{2}\right\rangle $
when we assume vacuum inputs at $\tau=0$. We have introduced constants
$g_{x}$, which can be chosen to minimise the variance relative to
the Heisenberg quantum noise level of the amplitudes $X_{A}$ and
$P_{A}$. We find this value of $g_{x}$ by standard procedures
\cite{mdrepr-1,rmp-1}:
\begin{equation}
\frac{\partial}{\partial g_{x}}\Delta^{2}\left(X_{A}-g_{x}X_{B}\right)=0,\qquad\Leftrightarrow g_{x}=\frac{\left\langle X_{A}X_{B}\right\rangle }{\left\langle X_{B}^{2}\right\rangle }\label{eq:g}
\end{equation}
For the system described by the Hamiltonian of Eq. (1), we find the
optimal value is $g_{x}=-\tanh2r$.  The optimal variance is given
by:
\begin{eqnarray}
\Delta^{2}\left(X_{A}-g_{x}X_{B}\right) & = & \left\langle X_{A}^{2}\right\rangle -\frac{\left\langle X_{A}X_{B}\right\rangle ^{2}}{\left\langle X_{B}^{2}\right\rangle }\label{eq:solcorre-2}
\end{eqnarray}
which for the parametric system (1) becomes $\Delta^{2}\left(X_{A}-g_{x}X_{B}\right)=\frac{1}{\cosh2r}$.
Similarly we evaluate the variance of $P_{A}+g_{p}P_{B}$ , which
we will denote as $\Delta^{2}\left(P_{A}+g_{p}P_{B}\right)$: The
variance minimizes when 
\[
\frac{\partial}{\partial g_{p}}\Delta^{2}\left(P_{A}+g_{p}P_{B}\right)=0,\qquad\Leftrightarrow g_{p}=-\frac{\left\langle P_{A}P_{B}\right\rangle }{\left\langle P_{B}^{2}\right\rangle }
\]
 This optimal value of $g_{p}$ for system (1) is $g_{p}=-\tanh2r.$
The optimal variance is given by
\begin{eqnarray}
\Delta^{2}\left(P_{A}+g_{p}P_{B}\right) & = & \left\langle P_{A}^{2}\right\rangle -\frac{\left\langle P_{A}P_{B}\right\rangle ^{2}}{\left\langle P_{B}^{2}\right\rangle }\label{eq:mom}
\end{eqnarray}
which becomes $\Delta^{2}\left(P_{A}+g_{p}P_{B}\right)=\frac{1}{\cosh2r}$
for Eq. (1).

\subsection{EPR steering correlations}

The criterion for the EPR paradox introduced in  \cite{mdrepr-1}
is also a criterion for EPR steering \cite{EPRsteering-1,hw-steering-1,hw2-steering-1}.
This EPR steering criterion is defined as the square root of the variance
product:
\begin{equation}
EPR_{A|B}(g_{x},g_{p})=\Delta\left(X_{A}-g_{x}X_{B}\right)\Delta\left(P_{A}+g_{p}P_{B}\right)\label{eq:epr5}
\end{equation}
We observe EPR steering (of system $A$ by $B$) whenever
\begin{equation}
EPR_{A|B}(g_{x},g_{p})<1\label{eq:ercrit}
\end{equation}
The ideal correlations created by the parametric down conversion Hamiltonian
interaction (1) are obtained in the limit of $r\rightarrow\infty$
and correspond to the EPR variances becoming zero, and hence $EPR\rightarrow0$. 

The EPR steering condition (\ref{eq:ercrit}) by its very definition
will negate all \emph{local hidden state }(LHS) models of the type
\cite{hw-steering-1,hw2-steering-1}
\begin{equation}
\langle X_{A}^{\theta}X_{B}^{\phi}\rangle\equiv\langle\hat{X}_{A}^{\theta}X_{B}^{\phi}\rangle=\int P(\lambda)\langle X_{A}^{\theta}\rangle_{\lambda}\langle X_{B}^{\phi}\rangle_{\lambda}d\lambda\label{eq:m4-1}
\end{equation}
These LHS models are similar to the local hidden variable models introduced
by Bell \cite{Bell,key-3}. Here $\lambda$ represents hidden variable
parameters and $P(\lambda)$ is the probability distribution for these
parameters. The $P(\lambda)$ is independent of the choice of measurement
($\theta$ and $\phi$), which is made \emph{after} the generation
and separation of the subsystems $A$ and $B$. In this model, the
$\langle X_{A}^{\theta}\rangle_{\lambda}$ is the average value for
the result $X_{A}^{\theta}$, given the hidden variable state specified
by $\lambda$. The $\langle X_{B}^{\phi}\rangle_{\lambda}$ is defined
similarly. For the LHS model however, there is the additional \emph{asymmetrical}
constraint that the\textcolor{black}{{} }\textcolor{black}{\emph{local
}}\textcolor{black}{hidden variable moments (such as $\langle X_{A}^{\theta}\rangle_{\lambda}$)
for system $A$ be consistent with measurements of some local observables
(for example position and momentum)}\textcolor{black}{\emph{ }}\textcolor{black}{at
$A$}\textcolor{black}{\emph{, }}\textcolor{black}{and is thus able
to be described as arising from a local}\textcolor{black}{\emph{ quantum}}\textcolor{black}{{}
density operator $\rho_{A}^{\lambda}$.}\textcolor{blue}{}

We can define the minimum value of $EPR$ after optimising $g_{x}$
and $g_{p}$ as: $EPR_{A|B}=min\{EPR_{A|B}(g_{x},g_{p})\}$. Thus,
for the Hamiltonian of Eq. (\ref{eq:Hmatwo-modesq}):
\begin{equation}
EPR_{A|B}=\frac{1}{\cosh2r}\label{eq:eprsoln}
\end{equation}
as derived originally in \cite{mdrepr-1}. Ideal EPR correlations
are obtained as $r\rightarrow\infty$, in which case $EPR_{A|B}\rightarrow0$,
and we say the ``steering is perfect''.

\subsection{Reservoir coupling}

We now examine the effect on EPR steering if the systems $A$ and
$B$ are independently coupled to heat bath reservoirs (Figure 1).
We assume that the parametric interaction $H$ given by (1) that generates
a two-mode squeezed state is turned off at the time $t=0$, and the
system left to decay. The two-mode squeezed state is a so-called Gaussian
state, meaning that its characteristic function is Gaussian \cite{gauss}.
Within the constraint of two-mode Gaussian states and measurements,
the condition (\ref{eq:ercrit}) will provide a \emph{necessary and
sufficient} test of steering \cite{hw-steering-1,hw2-steering-1}.
This makes the criterion valuable, for understanding the effects of
decoherence in Gaussian systems.

Thus, we consider a system \emph{prepared} in a two-mode squeezed
state at time $t=0$. In principle, the modes $a$ and $b$ can be
spatially separated. We consider the coupling of each mode $a$ and
$b$ to independent heat baths (reservoirs) with thermal occupation
numbers $n_{a}$ and $n_{b}$ respectively. The solutions after coupling
to the reservoir are straightforward to evaluate using the operator
Langevin equations that describe the evolution of the mode operators
\cite{crispn paper,key-21}:
\begin{eqnarray}
\dot{a} & = & -\gamma_{a}a{\color{red}{\color{black}{\color{black}}+}}\sqrt{2\gamma_{a}}\varGamma_{a}(t)\nonumber \\
\dot{b} & = & -\gamma_{b}b{\color{red}{\color{black}{\color{black}}+}}\sqrt{2\gamma_{b}}\varGamma_{b}(t)\label{eq:oplang}
\end{eqnarray}
Here $\gamma_{a}$ and $\gamma_{b}$ describe the decay rates (losses)
that are induced by the reservoirs. The quantum reservoir operators
$\Gamma(t)$ have nonzero correlations given by $\langle\Gamma_{a}^{\dagger}(t)\Gamma_{a}(t')\rangle=n_{a}\delta(t-t')$
and $\langle\Gamma_{a}(t)\Gamma_{a}^{\dagger}(t')\rangle=(n_{a}+1)\delta(t-t')$\textcolor{red}{{}
}\textcolor{black}{where} the numbers $n_{a}$ and $n_{b}$ give the
level of thermal occupation of the reservoirs. \textcolor{red}{}
Solutions are
\begin{eqnarray}
a(t) & = & e^{-\gamma_{a}t}a(0){\color{red}{\color{black}+}}\sqrt{2\gamma_{a}}\int_{0}^{t}e^{-\gamma_{a}(t-t^{\prime})}\Gamma_{a}\left(t^{\prime}\right)dt'\label{eq:soloplang}
\end{eqnarray}
From this, we can calculate the moments at a later time, in terms
of the initial moments:
\begin{eqnarray}
\langle a^{\dagger}(t)a(t)\rangle & = & e^{-2\gamma_{a}t}\langle a^{\dagger}(0)a(0)\rangle+n_{a}(1-e^{-2\gamma_{a}t})\nonumber \\
\langle a(t)b(t)\rangle & = & e^{-(\gamma_{a}+\gamma_{b})t}\langle a(0)b(0)\rangle\label{eq:res-1}
\end{eqnarray}
The solution for $\langle b^{\dagger}(t)b(t)\rangle$ is obtained
from that for $\langle a^{\dagger}(t)a(t)\rangle$, but exchanging
the letters $a$ with $b$. The initial moments are given by the solutions
found in Section II.A. The final moments after reservoir coupling
are\textcolor{red}{}
\begin{eqnarray}
\left\langle a^{\dagger}(t)a(t)\right\rangle  & = & n_{a}\left(1-e^{-2t\gamma_{a}}\right)+e^{-2t\gamma_{a}}\sinh^{2}\left(r\right)\nonumber \\
\left\langle a(t)b(t)\right\rangle  & = & -e^{-\left(\gamma_{a}+\gamma_{b}\right)t}\cosh r\sinh r\label{eq:7-1}
\end{eqnarray}
The covariance matrix $V$ is defined as:
\begin{equation}
V=\left[\begin{array}{cccc}
\left\langle X_{A}^{2}\right\rangle  & \left\langle X_{A}P_{A}\right\rangle  & \left\langle X_{A}X_{B}\right\rangle  & \left\langle X_{A}P_{B}\right\rangle \\
\left\langle P_{A}X_{A}\right\rangle  & \left\langle P_{A}^{2}\right\rangle  & \left\langle P_{A}X_{B}\right\rangle  & \left\langle P_{A}P_{B}\right\rangle \\
\left\langle X_{B}X_{A}\right\rangle  & \left\langle X_{B}P_{A}\right\rangle  & \left\langle X_{B}^{2}\right\rangle  & \left\langle X_{B}P_{B}\right\rangle \\
\left\langle P_{B}X_{A}\right\rangle  & \left\langle P_{B}P_{A}\right\rangle  & \left\langle P_{B}X_{B}\right\rangle  & \left\langle P_{B}^{2}\right\rangle 
\end{array}\right]\label{eq:v}
\end{equation}
Hence we find $V_{14}=V_{23}=0$, $V_{12}=V_{34}=i$ , $V_{11}=V_{22}$,
$V_{33}=V_{44}$, $V_{24}=-V_{13}$ where
\begin{eqnarray}
V_{11} & = & e^{-2\gamma_{a}t}\cosh2r+\left(1-e^{-2\gamma_{a}t}\right)\left(1+2n_{a}\right)\nonumber \\
V_{33} & = & e^{-2\gamma_{b}t}\cosh2r+\left(1-e^{-2\gamma_{b}t}\right)\left(1+2n_{b}\right)\nonumber \\
V_{24} & = & e^{-\left(\gamma_{a}+\gamma_{b}\right)t}\sinh2r\label{eq:13}
\end{eqnarray}
 Applying the results of the Section II.A, we calculate that 

\begin{widetext}
\begin{equation}
\footnotesize
EPR_{A|B}=\frac{\cosh2r\left[e^{-2\gamma_{a}t}\left(1-e^{-2\gamma_{b}t}\right)\left(1+2n_{b}\right)+e^{-2\gamma_{b}t}\left(1-e^{-2\gamma_{a}t}\right)\left(1+2n_{a}\right)\right]+e^{-2(\gamma_{a}+\gamma_{b})t}+\left(1+2n_{a}\right)\left(1+2n_{b}\right)\left(1-e^{-2\gamma_{b}t}\right)\left(1-e^{-2\gamma_{a}t}\right)}{e^{-2\gamma_{b}t}\cosh2r+\left(1-e^{-2\gamma_{b}t}\right)\left(1+2n_{b}\right)}
\end{equation}
\end{widetext}

\begin{figure}
\includegraphics[width=0.5\columnwidth]{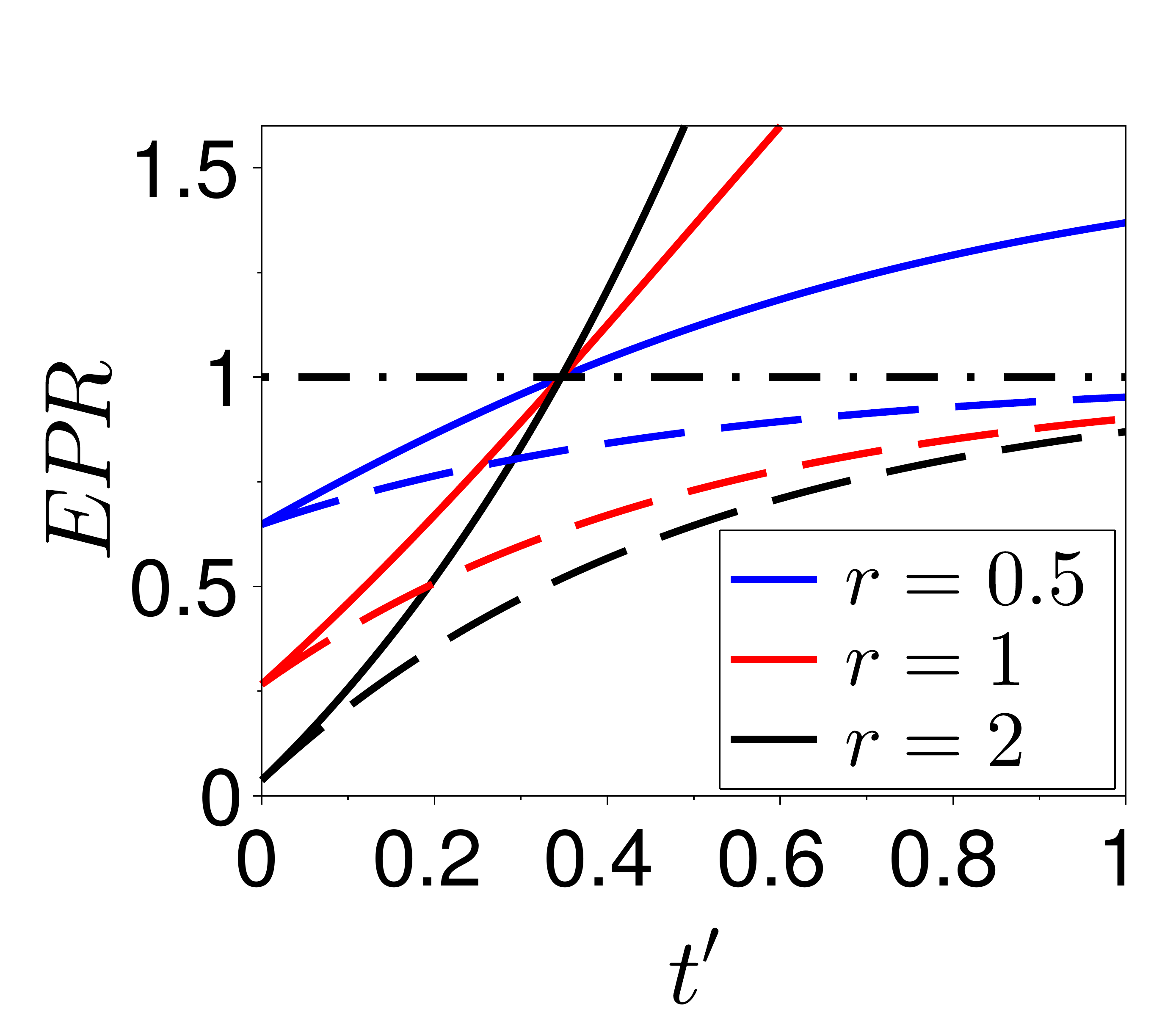}\includegraphics[width=0.5\columnwidth]{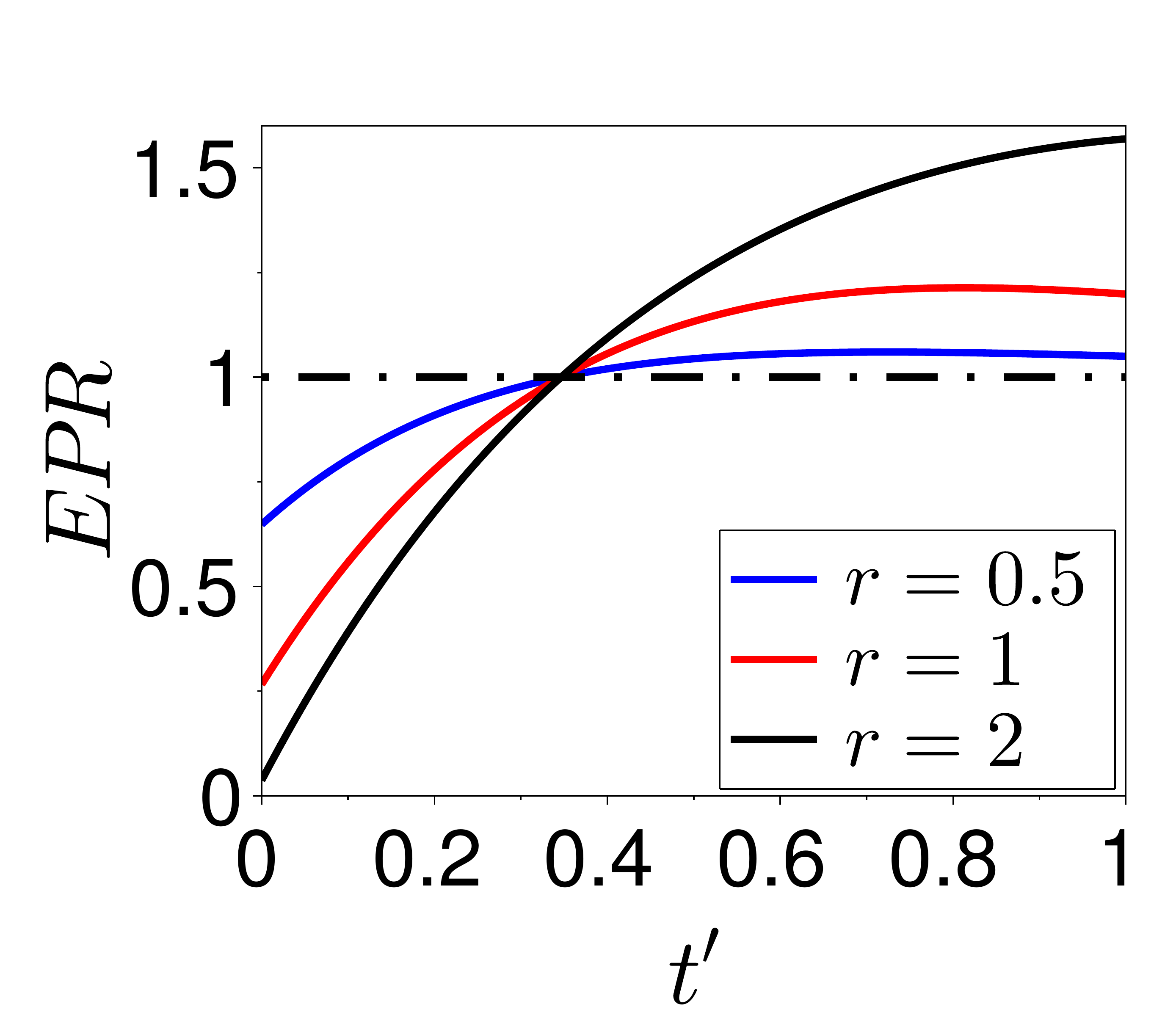}

\protect\caption{\emph{\label{fig:observables-1000}}The decoherence of two-mode EPR
steering with no thermal excitation ($n_{a}=n_{b}=0$). A sudden-death
type of decoherence is noted when it is the system ``doing the steering''
that is is coupled to the reservoir. The steering signatures $EPR_{A|B}$
and $EPR_{B|A}$ are plotted versus $t'=\gamma_{b}t$, for various
values of squeeze parameter $r$. EPR steering is obtained for the
Gaussian system iff $EPR<1$\textbf{ }and the strongest steering is
when\textbf{ $EPR\rightarrow0$. Left graph:} Reservoir coupling to
system $B$ only ($\gamma_{a}=0$). Solid lines correspond to $EPR_{A\vert B}$
while dashed lines correspond to $EPR_{B\vert A}$.\textbf{ }Curves
are (bottom to top on the far left) $r=2$ (black), $r=1$ (red),
$r=0.5$ (blue). \textbf{Right graph:} Plots of $EPR_{A|B}$ for symmetric
coupling to the reservoirs: $\gamma_{a}=\gamma_{b}$\textcolor{blue}{.
}\textcolor{black}{In this case, $EPR_{A|B}=EPR_{B|A}$}\textcolor{blue}{}\textcolor{black}{.}\textcolor{red}{}}
\end{figure}

\subsection{Decoherence of EPR steering with no thermal noise}

We first assume negligible thermal noises: we put $n_{a}=n_{b}=0$
in the solution given by Eq. (17). The decoherence effect on the system
by the reservoirs manifests as losses, parametrised by $\gamma_{a}$
and $\gamma_{b}$. We will study the effect on the steering parameters,
showing the effect of loss to be asymmetrical with respect to the
two systems $A$ and $B$.

\subsubsection{\textcolor{black}{Damping the steering system: Steering sudden-death}}

\textcolor{black}{Let us suppose that the reservoirs are coupled }\textcolor{black}{\emph{asymmetrically}}\textcolor{black}{,
so that $\gamma_{a}\rightarrow0$ i.e. only the mode $b$ is coupled
to a reservoir. In this case, we find for the steering of $A$ (mode
$a$)
\begin{eqnarray}
EPR_{A\vert B} & = & \frac{e^{-2\gamma_{b}t}+\cosh2r\left[1-e^{-2\gamma_{b}t}\right]}{1+e^{-2\gamma_{b}t}\left(\cosh2r-1\right)}
\end{eqnarray}
}The plots of Figure 2 (left graph) indicate that where the losses
are entirely on the steering mode $B$, the decoherence of EPR steering
is substantial. After a time given by $\gamma_{b}t=\frac{\ln2}{2}$,
\textcolor{red}{}steering (as measured by $EPR_{A|B}$) in that direction
is lost. The loss is inevitable, regardless of how much steering is
present in the initial two-mode quantum system, as shown by the results
for the different values of squeeze parameter $r$: We also note that
the\emph{ cut-off time} for steering is independent of the amount
of steering $r$ in the initial two-mode system. The behavior observed
here is analogous to the ``entanglement sudden-death'', that has
been observed for the decoherence of entanglement \cite{eberly,key-23,key-22}.

\subsubsection{Damping the steered system}

\textcolor{black}{However, the sudden-death effect is not apparent
when the ``steered'' system is lossy. We again assume $\gamma_{a}\rightarrow0$
and evaluate the steering of the damped system, by the undamped system:}
\begin{equation}
EPR_{B\vert A}=\frac{e^{-2\gamma_{b}t}+\cosh2r\left[1-e^{-2\gamma_{b}t}\right]}{\cosh2r}\label{eq:two}
\end{equation}
The plots of Figure 2 (left graph) reveal the nature of the decoherence
when only the steered system is damped. Here, we see that there is
much less sensitivity of the steering to the losses. In fact, while
the EPR steering is certainly diminished by the dissipation, there
is no cut-off, or sudden death, but rather steering is still possible
for arbitrary times, $t\rightarrow\infty$. We remark that the increased
sensitivity to losses affecting the steering system, as compared to
the steered system, has been noted in earlier papers \cite{rmp-1,onewaysteer,steer thmurray,monog}.

\subsubsection{Symmetric decoherence}

We next consider a reservoir with symmetric damping $\gamma_{a}=\gamma_{b}$.
\begin{eqnarray}
EPR_{A|B} & = & \frac{e^{-4\gamma_{b}t}+\left(1-e^{-2\gamma_{b}t}\right)^{2}}{1+e^{-2\gamma_{b}t}\left(\cosh2r-1\right)}\nonumber \\
 & + & \frac{2\cosh2r\left[e^{-2\gamma_{b}t}-e^{-4\gamma_{b}t}\right]}{1+e^{-2\gamma_{b}t}\left(\cosh2r-1\right)}\label{eq:sym}
\end{eqnarray}
The results of Figure 2 (right graph) show, as might be expec\textcolor{black}{ted,
the effect is dominated by the fact that the steering system is damped.
Here steering }(as measured by $EPR_{A|B}$)\textcolor{black}{{} is
also lost at $\gamma_{b}t=\frac{\ln2}{2}$. }

\subsection{Decoherence of EPR steering with thermal noise}

Now we consider the more complex interaction where there is additional
thermal noise for each reservoir. \textcolor{red}{}Here we use the
full expressions given by Eq. (17). Our results in Figures 3 and 4
reveal several features. We note that the loss of steering is rapid
and complete (``sudden-death'') if the thermal noise is placed on
the system that is being steered. This effect has been noticed in
previous studies of thermal steering and is especially important for
asymmetrical systems such as found in opto-mechanics \cite{simon,thermal he andR,karenbec,thermal atomic ensembles}.
The effect is more significant as the thermal noise increases, but
is reduced for higher $r$ (Figure 4).
\begin{figure}

\includegraphics[scale=0.2]{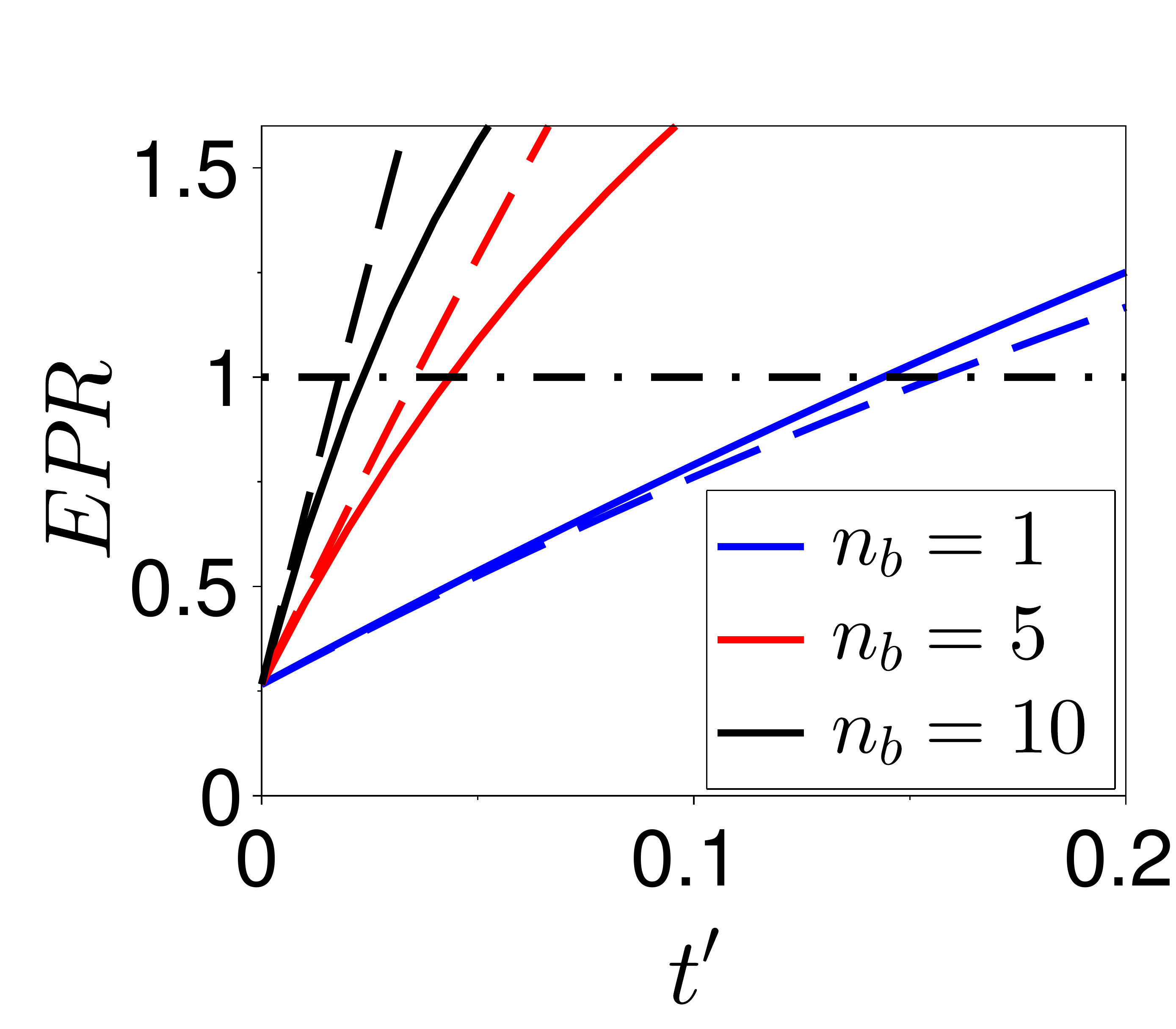}

\protect\caption{\emph{\label{fig:EPR}}The decoherence of two-mode EPR steering with
a thermal reservoir coupled to system $B$ only. Here $\gamma_{a}=0$,
$r=1$. We plot the steering signature versus $t'=\gamma_{b}t$. The
thermal noise creates a sudden-death type decoherence effect when
it is the system that is ``being steered'' that is coupled to the
reservoir. Solid lines correspond to $EPR_{A\vert B}$ while dashed
lines correspond to $EPR_{B\vert A}$. EPR-steering is obtained when
$EPR<1$ and the strongest steering is for\textbf{ $EPR\rightarrow0$.}
We plot different values of $n_{b}$: from bottom to top (for each
line type), $n_{b}=1$ (blue), $n_{b}=5$ (red), $n_{b}=10$ (black).
\textcolor{blue}{}}
\end{figure}
\begin{figure}
\begin{centering}

\par\end{centering}

\begin{centering}
\includegraphics[width=0.5\columnwidth]{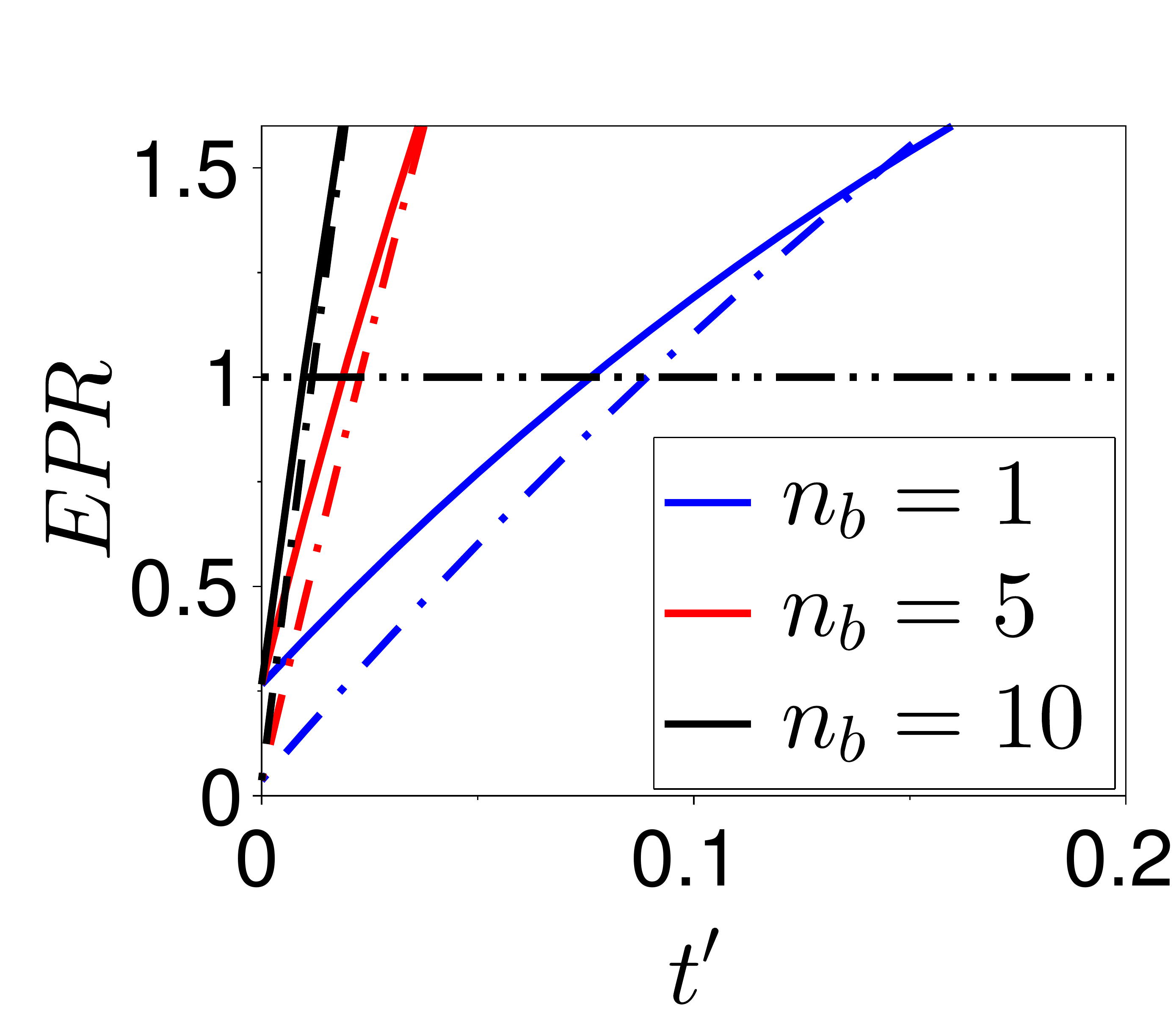}\includegraphics[width=0.5\columnwidth]{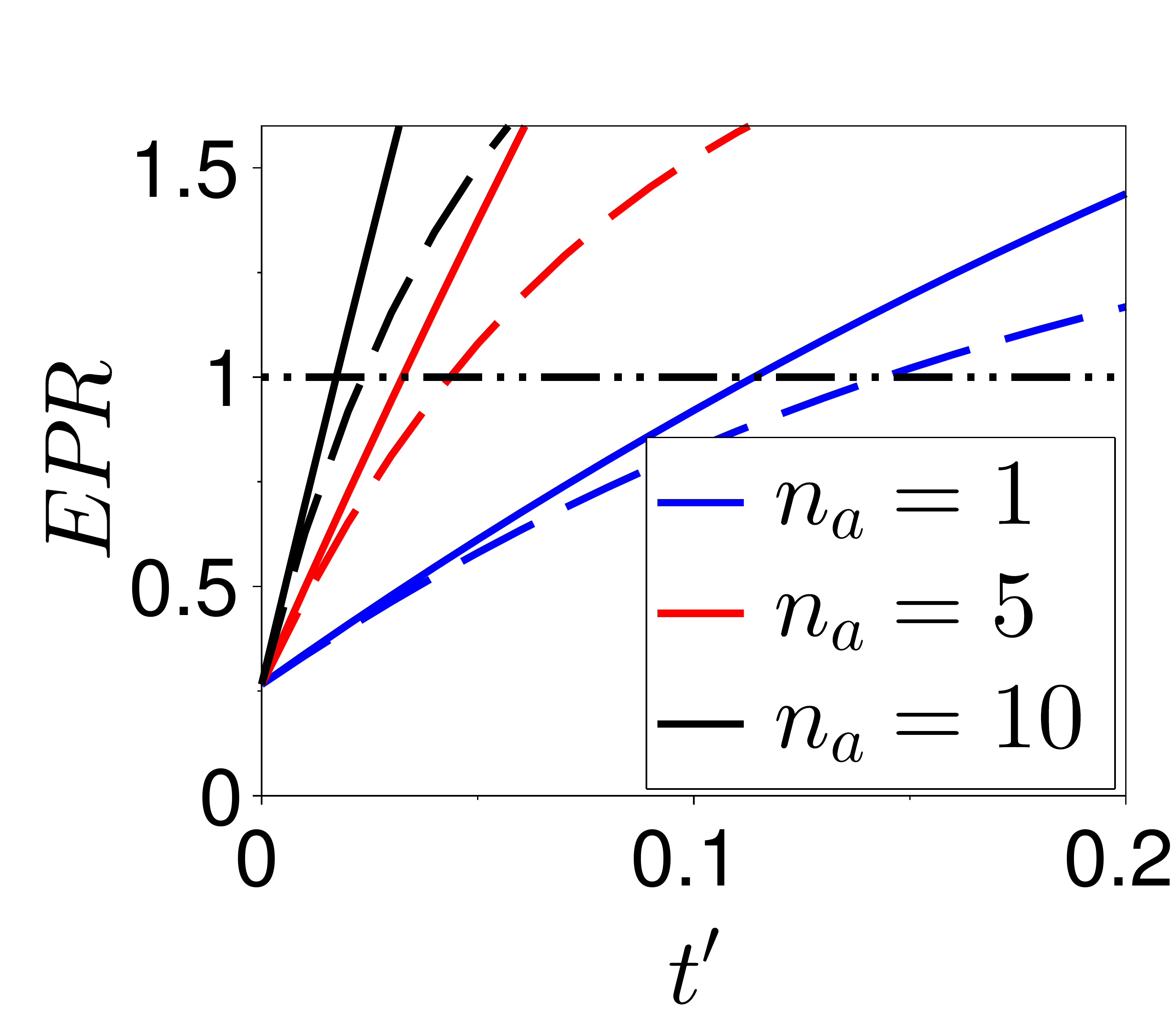}
\par\end{centering}

\begin{centering}

\par\end{centering}

\protect\caption{\emph{\label{fig:symmetric}}The decoherence of two-mode EPR steering
with thermal reservoirs. EPR-steering is obtained when $EPR<1$ and
the strongest steering is for\textbf{ $EPR\rightarrow0$.} \textbf{Left
graph:} Reservoirs coupled to both systems $A$ and $B$ symmetrically.
We plot $EPR_{A|B}$ vs $t^{\prime}=\gamma_{b}t$. Here $\gamma_{a}=\gamma_{b}$,
$n_{a}=n_{b}$, $r=1$. We plot different values of $n_{b}$, given
from bottom to top (for each line type) , $n_{b}=1$ (blue), $n_{b}=5$
(red), $n_{b}=10$ (black). \textcolor{black}{Solid lines correspond
to $r=1$ while dashed-dotted lines correspond to $r=2$. }\textbf{Right}
\textbf{graph}: Thermal reservoir coupling to system $A$ and cold
reservoir coupling to $B$ ($\gamma_{a}=\gamma_{b}$, $n_{a}=1,5,10$).
Solid lines correspond to $EPR_{A\vert B}$ while dashed lines correspond
to $EPR_{B\vert A}$.}
\end{figure}

\subsection{Comparing with the decoherence of entanglement}

\begin{figure}

\begin{centering}

\par\end{centering}

\begin{centering}
\includegraphics[width=0.5\columnwidth]{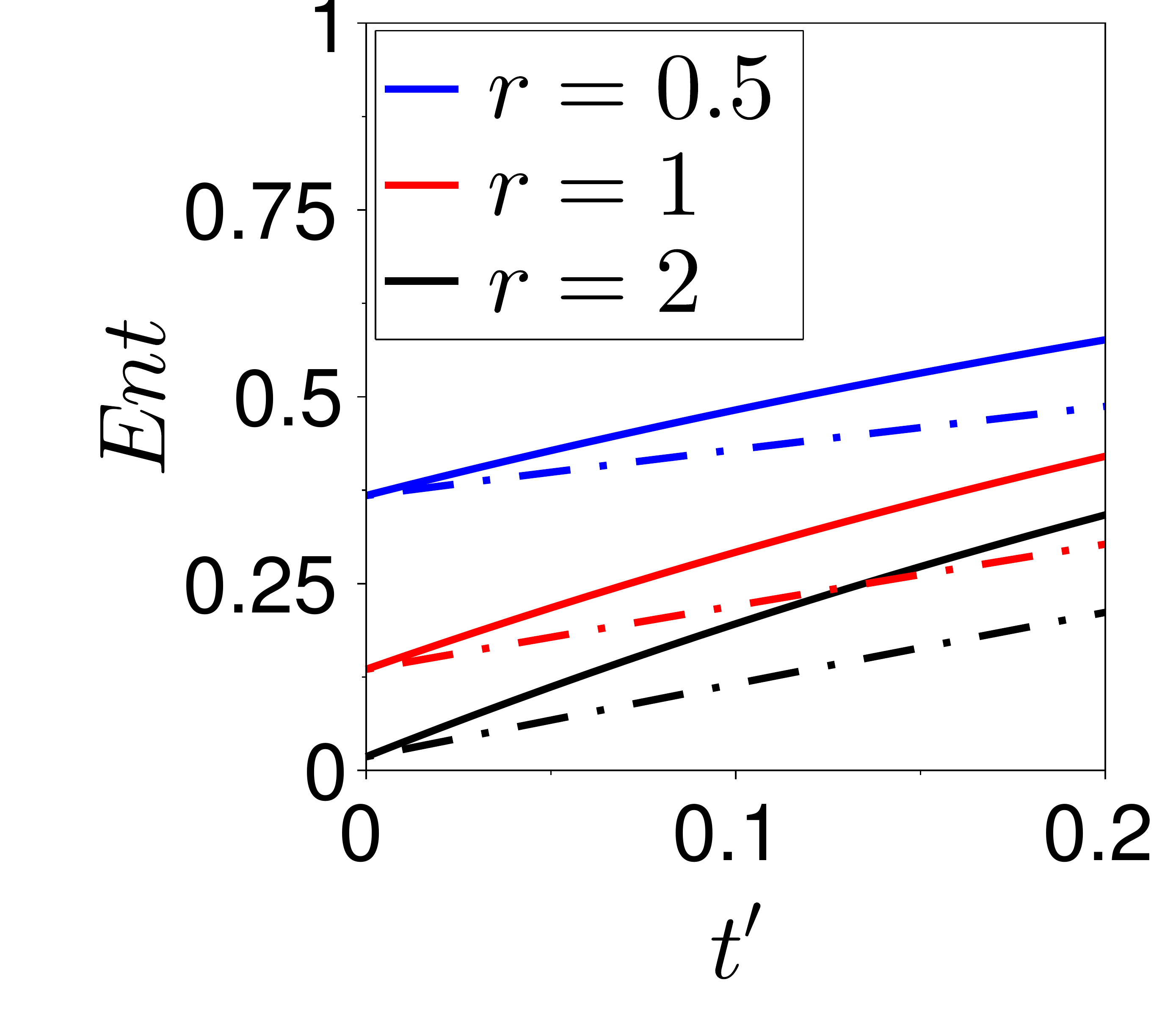}\includegraphics[width=0.5\columnwidth]{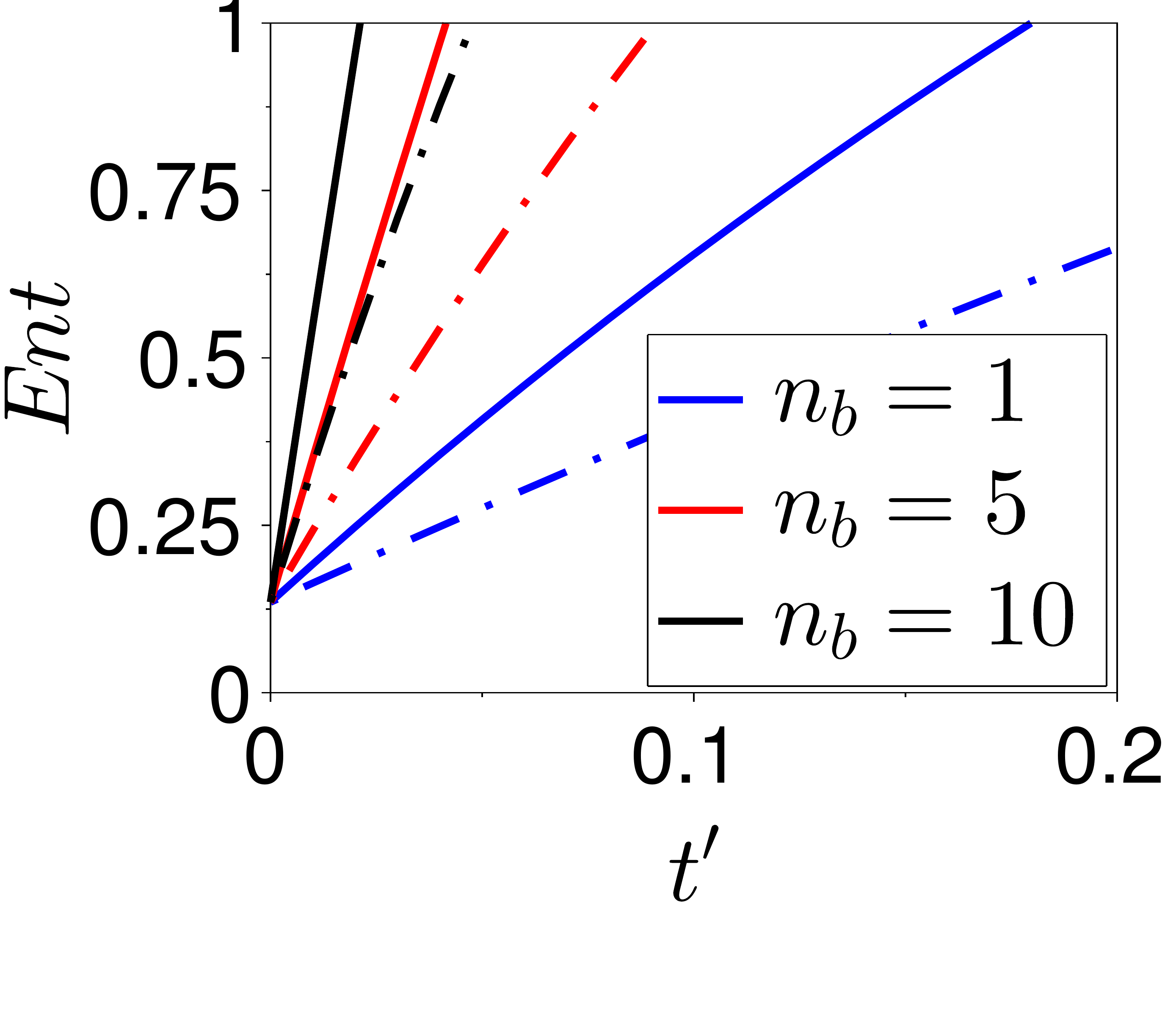}
\par\end{centering}

\begin{centering}

\par\end{centering}

\protect\caption{\emph{\label{fig:observables-1000-4}}The decoherence of the entanglement
of the two-mode squeezed state under the action of a reservoir. The
entanglement signature for Gaussian systems $Ent$ is plotted versus
scaled time $t'$, under different sorts of reservoir coupling. Entanglement
is obtained iff $Ent<1$ and stronger entanglement corresponds to
$Ent\rightarrow0$. \textbf{Left graph:} Here we take all thermal
noise zero ($n_{a}=n_{b}=0$). Solid curves show the symmetric case
$\gamma_{a}=\gamma_{b}$. Dashed-dotted curves are for the asymmetric
case, for a reservoir coupling to system $B$ only ($\gamma_{a}=0$)
when $t'=\gamma_{b}t$; or, for a reservoir coupling to system $A$
only ($\gamma_{b}=0$) when $t'=\gamma_{a}t$. We plot different values
of $r$, given from bottom to top (for each line type): $r=2$ (black),
$r=1$ (red), $r=0.5$ (blue). \textbf{Right graph:} Here we study
thermal reservoirs. Solid curves show the symmetric case $n_{a}=n_{b}$
and $\gamma_{a}=\gamma_{b}$. Dashed-dotted curves are for the asymmetric
case, for a reservoir coupling to system $B$ only ($\gamma_{a}=0$)
when $t'=\gamma_{b}t$. We plot different values of thermal noise,
given from bottom to top (for each line type) $n_{b}=1$ (blue), $n_{b}=5$
(red), $n_{b}=10$ (black).\textcolor{red}{}}
\end{figure}

The effects of the reservoir on the steering are asymmetrical. This
will not be true for entanglement: Entanglement is symmetrically defined,
with respect to the two different systems involved, and the numerical
value we assign to the entanglement will be unchanged under the exchange
$A\longleftrightarrow B$. The relationships between entanglement
and EPR steering for Gaussian states have been recently studied \cite{prlgaussadesso,key-24}.
For Gaussian states, it is possible to give a quantification of steering
\cite{prlgaussadesso,hw-steering-1,hw2-steering-1,key-24}, which
is an asymmetrical quantum correlation, similar to quantum discord
\cite{prlstefano,key-32,key-31}.

We may evaluate the entanglement via the parameter $Ent=\Delta(X_{A}-gX_{B})\Delta(P_{A}+gP_{B})/\{1+g^{2}\}$
\cite{itent,key-33}. The inequality $Ent<1$ is a necessary and sufficient
condition for entanglement of two-mode Gaussian systems for the case
we examine here where $\Delta\left(X_{A}-gX_{B}\right)=\Delta\left(P_{A}+gP_{B}\right)$.
The condition for optimally chosen $g$ has been shown equivalent
to the Peres-Simon condition in that case \cite{key-24}. Thus, in
this section, we consider the entanglement parameter
\begin{equation}
Ent=\frac{\Delta(X_{A}-gX_{B})\Delta(P_{A}+gP_{B})}{1+g^{2}}\label{eq:syment}
\end{equation}
 First we find the value of $g$ that minimizes the parameter $Ent.$
We evaluate $\frac{\partial Ent}{\partial g}$, in this case we know
that $\Delta\left(X_{A}-gX_{B}\right)=\Delta\left(P_{A}+gP_{B}\right)$
hence:
\begin{eqnarray}
\frac{\partial Ent}{\partial g} & = & \frac{\partial}{\partial g}\frac{\Delta^{2}\left(X_{A}-gX_{B}\right)}{1+g^{2}}\nonumber \\
 & = & \frac{-2\left\langle X_{A}X_{B}\right\rangle +2g\left(\left\langle X_{B}^{2}\right\rangle -\left\langle X_{A}^{2}\right\rangle \right)+2g^{2}\left\langle X_{A}X_{B}\right\rangle }{\left(1+g^{2}\right)^{2}}\nonumber \\
\label{eq:delgent}
\end{eqnarray}
Next $\frac{\partial Ent}{\partial g}=0,$ 
\[
\Leftrightarrow-\left\langle X_{A}X_{B}\right\rangle +g\left(\left\langle X_{B}^{2}\right\rangle -\left\langle X_{A}^{2}\right\rangle \right)+g^{2}\left\langle X_{A}X_{B}\right\rangle =0
\]
Hence the value of $g$ that minimizes $Ent$ is \cite{key-24}:
\begin{eqnarray}
g & = & \frac{-\left(\left\langle X_{B}^{2}\right\rangle -\left\langle X_{A}^{2}\right\rangle \right)+\sqrt{\left(\left\langle X_{B}^{2}\right\rangle -\left\langle X_{A}^{2}\right\rangle \right)^{2}+4\left\langle X_{A}X_{B}\right\rangle ^{2}}}{2\left\langle X_{A}X_{B}\right\rangle }\nonumber \\
\label{eq:gquad}
\end{eqnarray}
We note that if $\left\langle X_{A}^{2}\right\rangle =\left\langle X_{B}^{2}\right\rangle $,
then the value of $g$ that minimises the expression is $g=1$, but
where we have asymmetric reservoir effects, like different reservoir
couplings or thermal noises, the optimal value will be different to
$1$. Here we will use from Section II.A that:
\begin{eqnarray}
\left\langle X_{A}^{2}\right\rangle  & = & e^{-2\gamma_{a}t}\cosh2r+\left(1-e^{-2\gamma_{a}t}\right)\left(1+2n_{a}\right)\nonumber \\
\left\langle X_{A}X_{B}\right\rangle  & = & -e^{-\left(\gamma_{a}+\gamma_{b}\right)t}\sinh2r\nonumber \\
\left\langle X_{B}^{2}\right\rangle  & = & e^{-2\gamma_{b}t}\cosh2r+\left(1-e^{-2\gamma_{b}t}\right)\left(1+2n_{b}\right)\nonumber \\
\label{eq:4es}
\end{eqnarray}

Figure 5 plots the entanglement $Ent$ for various reservoir couplings.
We see from Figure 5 (left graph) that where there is no thermal noise,
the entanglement decays steadily, but is never destroyed completely.
This effect was noted in \cite{buono bow,key-34}. The larger values
of $r$ correspond to larger amounts of entanglement for the initial
state before decoherence, and we note that while the decay is sharper
for higher $r$, a larger initial amount of EPR entanglement will
ensure larger EPR entanglement for all later times.\textcolor{red}{{}
}\textcolor{black}{The Figure 5 (right graph) shows that when the
reservoirs are thermally excited, the entanglement is totally destroyed
(in sudden-death fashion) after a finite time. This time is shortened,
by increasing the amount of thermal excitation of the reservoir (on
either system).}

\section{Decoherence of the steerability of a S-cat state}

Traditionally, most studies of decoherence have centred around the
Schrodinger cat \cite{legdecoh,yurke stoler,key-12,key-13,key-14}.
We noticed in the study of the decoherence of steering for the two-mode
squeezed state that for stronger EPR effects, the decay was brought
about more sharply, to give overall decoherence times of a similar
order (Figure 2). This is consistent with the overall intuition about
decoherence in quantum mechanics, that it acts to destroy the more
extreme (larger) effects more quickly so that they are not observed
in nature. This has been studied for the Schrodinger cat example,
where one considers a system initially prepared in superposition of
two states macroscopically distinguishable e.g. in phase space \cite{legdecoh,cats,cats3,yurke stoler,key-12,key-13,key-14}.
The coupling to a reservoir induces a decay of the superposition.
Where the separation in phase space increases, calculation and experiments
show that the decay rate will increase. Thus, there is an explanation
of the transition from microscopic to macroscopic quantum mechanics.
Here, motivated by this, we examine the decoherence of the EPR steering
of a Schrodinger cat state.\textcolor{blue}{}

\subsection{Steering signature for a Schrodinger cat}

Consider the state:
\begin{equation}
|\psi\rangle=\frac{1}{\sqrt{2}}\{|-\alpha\rangle_{A}|\uparrow\rangle_{B}+e^{i\theta}|\alpha\rangle_{A}|\downarrow\rangle_{B}\}\label{eq:macrosup-1}
\end{equation}
which is the entangled Schrodinger cat state. We select $\theta=\pi/2$
and $\alpha$ to be real. Here, $|\alpha\rangle$ is the coherent
state for mode $a$ of system $A$ and $|\uparrow\rangle$, $|\downarrow\rangle$
are the eigenstates of the Pauli spin $\sigma_{z}$ of system $B$.
This state is similar to that described in Schrodinger's original
gedanken experiment \cite{schcat}, where a microscopic system becomes
entangled with a macroscopic one, the ``cat''. The spin-mode entangled
state given by Eq. (\ref{eq:macrosup-1}) has been the subject of
several experiments \cite{cats}. We first explain how one can signify
the Schrodinger cat, using EPR steering.  

\begin{figure}
\begin{centering}

\par\end{centering}

\begin{centering}
\includegraphics[width=0.5\columnwidth]{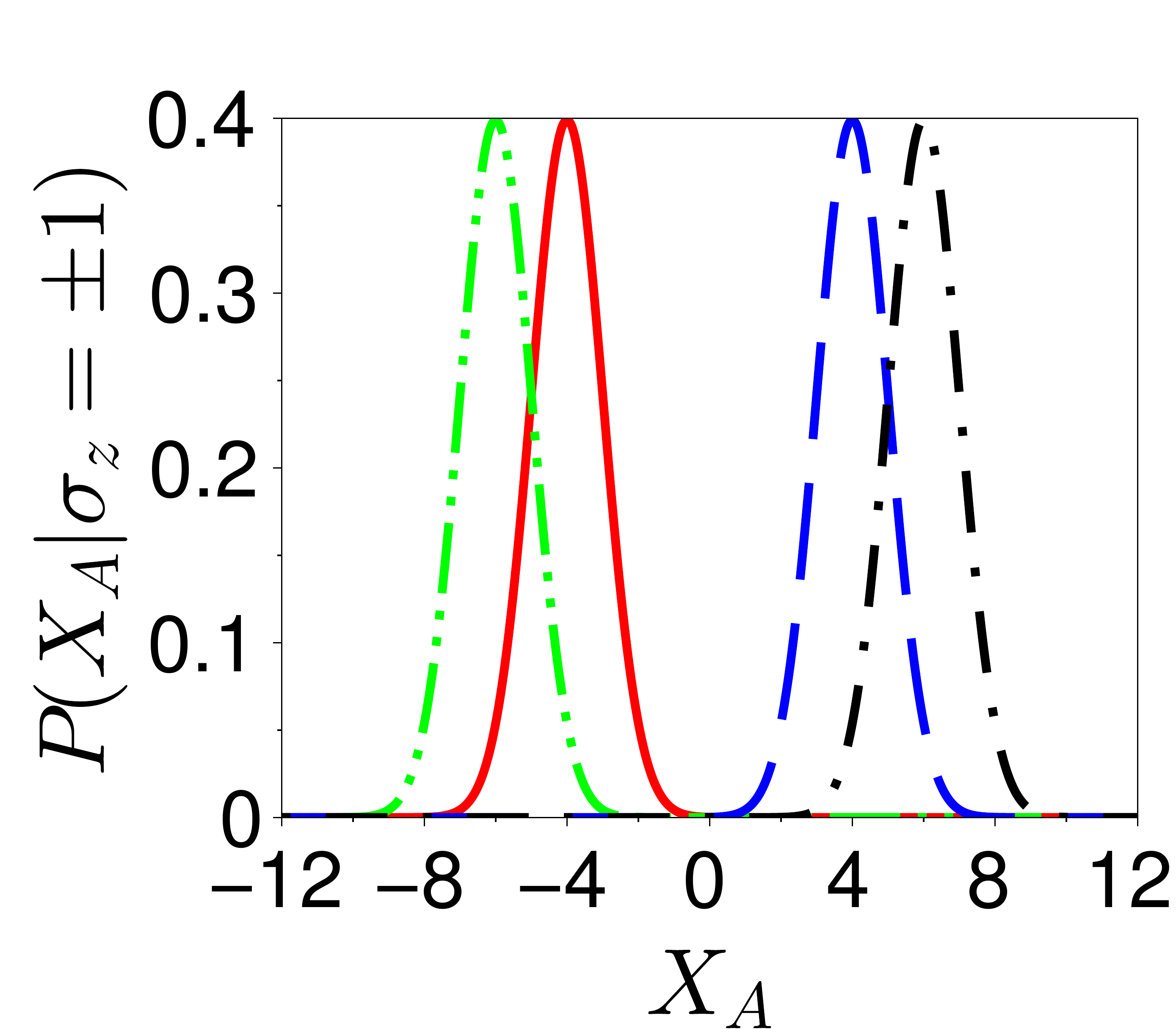}\includegraphics[width=0.5\columnwidth]{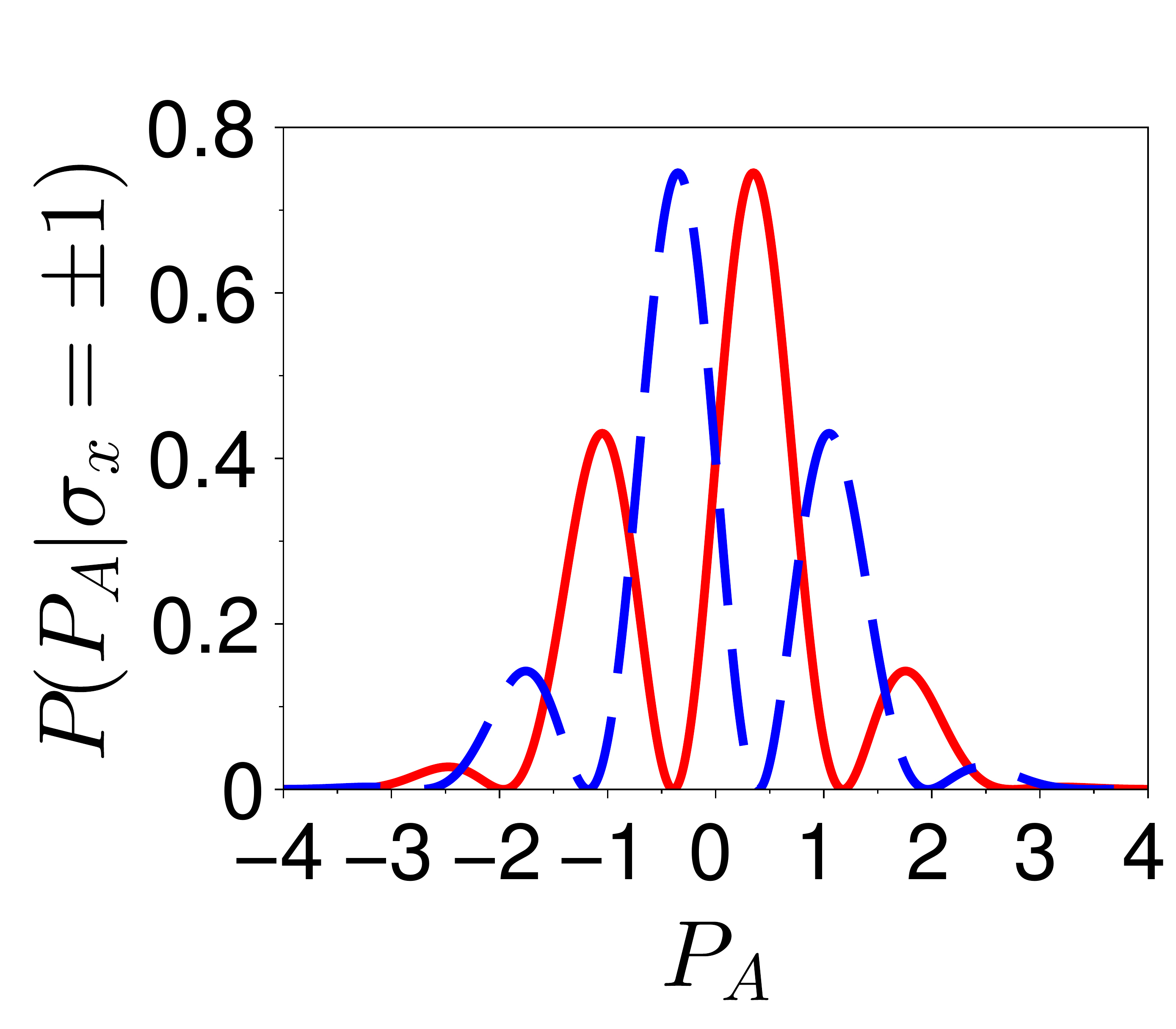}
\par\end{centering}

\protect\caption{\emph{\label{fig:PX}}The conditional distributions associated with
an EPR steering signature for the Schrodinger cat. \textbf{Left Graph:}
Conditional probability\textcolor{black}{{} for $X_{A}$ }given that
the result of a measurement $\sigma_{z}$ is $+1$ (red solid curve
is for $\alpha=2$, green dash-dotted curve is for $\alpha=3$) or
$-1$ (blue long dashed curve is for $\alpha=2$, black short dashed
curve is for $\alpha=3$).\textcolor{red}{{} }\textbf{\textcolor{black}{Right
Graph:}}\textcolor{black}{{} }Conditional probability for $P_{A}$ given
that the result of the measurement $\sigma_{x}$ is $+1$ (red solid
curve) or $-1$ (blue long dashed curve). Here $\alpha=2$.}
\end{figure}

We will derive the probability distributions $P(X_{A})$ (or $P(P_{A})$)
for the system $A$, given that a measurement of the Pauli spin $\sigma_{z}$
(or $\sigma_{x}$) at $B$ has been carried out to give a result $+1$
or $-1$.\textcolor{red}{}\textcolor{blue}{{} }\textcolor{red}{}
We define the scaled amplitudes for position and momentum by $a=\frac{1}{c}(\hat{x}+i\hat{p})$
where $a$ is the boson operator, and note that for real position
and momentum, $c=\sqrt{2}$. In terms of the quadrature phase amplitudes
$X_{A/B}$ and $P_{A/B}$ defined in Section II, however, we chose
the scaling $c=2$. We will distinguish the two cases by using lower
and upper case, respectively, and note we have dropped the use of
the operator hats where the meaning is clear, or to denote the outcomes
of the measurements. As expected from direct examination of the state
(\ref{eq:macrosup-1}), the distributions $P(x\bigl|+1)$ and $P(x\bigl|-1)$
for the result $x$ at $A$ given the outcomes $+1$ or $-1$ for
$\sigma_{z}$ at $B$, are the two Gaussian hills (Figure 6, left
graph): Specifically, \textcolor{red}{}\textbf{\textcolor{red}{}}\textcolor{black}{
\begin{eqnarray}
P\left(x\Bigl|\sigma_{z}=\pm1\right) & = & \left(\frac{2}{\pi}\right)^{\frac{1}{2}}\frac{1}{c}\exp\left[-2\frac{x^{2}}{c^{2}}-2\alpha^{2}\mp4\frac{x\alpha}{c}\right]\nonumber \\
\label{eq:scat2}
\end{eqnarray}
}

Next, we derive the conditional distributions for $P(P_{A})$ given
that a measurement of Pauli spin $\sigma_{x}$ at $B$ yields the
result $+1$ or $-1$. To evaluate this, we rewrite the state in
terms of the basis states $|\uparrow\rangle_{x}$ and $|\downarrow\rangle_{x}$
for the spin $\sigma_{x}$.\textcolor{blue}{}\textcolor{red}{}
\begin{eqnarray}
|\psi\rangle & = & \frac{1}{2}(|-\alpha\rangle_{A}+i|\alpha\rangle_{A})|\uparrow\rangle_{x,B}\nonumber \\
 & + & \frac{1}{2}(i|\alpha\rangle_{A}-|-\alpha\rangle_{A})|\downarrow\rangle_{x,B}\label{eq:ScXbasis-3}
\end{eqnarray}
At this point, we note the for other choices of $\theta$, normalisation
factors are more complicated. We find\textcolor{red}{{} }\textcolor{black}{that
the probability for the momentum given the results for spin $\sigma_{x}$
is}\textcolor{red}{{} }\textcolor{blue}{}\textcolor{red}{}
\begin{eqnarray}
P\left(p\Bigl|\sigma_{x}=\pm1\right) & = & \left(\frac{2}{\pi}\right)^{\frac{1}{2}}\frac{1}{c}e^{-2\frac{p^{2}}{c^{2}}}\left(1\pm\sin\left(4\frac{p\alpha}{c}\right)\right)\nonumber \\
\label{eq:fringe5}
\end{eqnarray}

\begin{figure}
\begin{centering}

\par\end{centering}

\begin{centering}

\par\end{centering}

\begin{centering}
\includegraphics[width=0.75\columnwidth]{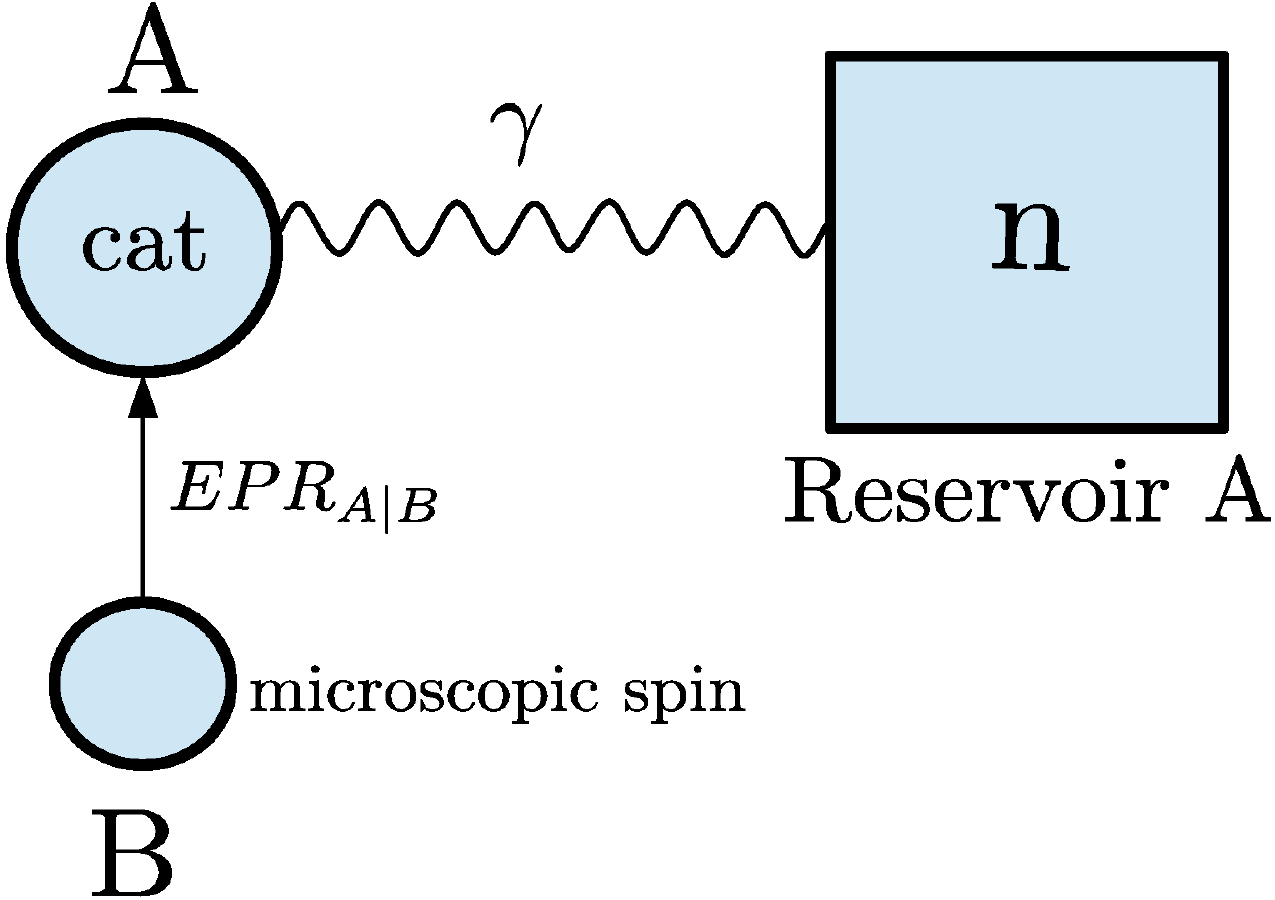}
\par\end{centering}

\protect\caption{\emph{\label{fig:cat-1}}The steering of a Schrodinger cat. We have
proposed a signature for the cat, based on EPR steering parameter
$EPR_{A|B}$. Here, measurements made on the microscopic spin system
$B$ ``steer'' the macroscopic system (the cat) $A$. The ``cat''
in this case is modelled as the superposition involving two coherent
states, $|\alpha\rangle$ and $|-\alpha\rangle$. The value of $\alpha$
determines the size of the Schrodinger cat. We consider that the ``cat''
being macroscopic is coupled to a thermal reservoir. We examine the
decay of the steering signature, with respect to the parameters of
the reservoir coupling.\textcolor{red}{}\textcolor{blue}{}}
\end{figure}

This allows us to calculate the conditional variances that give us
a signature for $EPR$ steering. An EPR steering of the ``cat''
is observed when\textcolor{blue}{}\textcolor{black}{
\begin{equation}
{\color{black}EPR=Var(X_{A}\Bigl|\sigma_{z})Var(P_{A}\Bigl|\sigma_{x})<1}\label{eq:s}
\end{equation}
}where here
\[
Var(X_{A}\Bigl|\sigma_{z})=P_{z}(+1)Var(X_{A}\Bigl|+1)+P_{z}(-1)Var(X_{A}\Bigl|-1)
\]
denotes the conditional variance for $X_{A}$ averaged over the outcomes
of $\sigma_{z}$. We have used the notation $Var(X_{A}\Bigl|\pm1)$
to mean the variance of $P(X_{A}\Bigl|\pm1)$, which is the probability
distribution for $X_{A}$ given the outcome $\pm1$ for the measurement
$\sigma_{z}$, respectively. The $Var\left(P_{A}|\sigma_{x}\right)$
is defined similarly, but for outcomes of $\sigma_{x}$. The right-side
bound is the quantum noise limit, as determined by the Heisenberg
uncertainty relation which in this case for the choice $c=2$ is $\Delta X_{B}\Delta P_{B}\geq1$.
We note that the derivation of the inequality as a signature of EPR
steering has been given in the Refs. \cite{EPRsteering-1,mdrepr-1,rmp-1}. 

The EPR quantity (\ref{eq:s}) can be evaluated: We see that for the
two Gaussian hills, the variance is at the quantum noise level (Figure
6, left graph): $Var(X_{A}\Bigl|\sigma_{z})=1$. \textcolor{red}{}On
the other hand, the variance associated with the momentum distributions
is reduced below $1$ (Figure 6, right graph), implying that $EPR_{A|B}<1$.
We can calculate the variance from the conditional distributions.
Alternatively, noting that the collapsed state for system $B$ given
a measurement $\sigma_{x}$ is the superposition $\frac{1}{2}(|-\alpha\rangle_{A}+i|\alpha\rangle_{A})$
(for result $+1$) or $\frac{1}{2}(i|\alpha\rangle_{A}-|-\alpha\rangle_{A})$
(for result $-1$), \textcolor{red}{} it is easy to use the methods
and results of the next section, to find that \textcolor{blue}{}
\begin{equation}
Var(P_{A}\Bigl|\sigma_{x})=1-4\alpha^{2}e^{-4\left|\alpha\right|^{2}}\label{eq:VarP_SECIIIA}
\end{equation}

The steering inequality (\ref{eq:s}) has been suggested in Ref. \cite{erci cav reid pra },
as a way to realise an EPR paradox with the Schrodinger cat state.
The steering cannot be obtained if the system is in the \emph{mixture
of states}, $|-\alpha\rangle_{A}|\uparrow\rangle_{B}$ and $|\alpha\rangle_{A}|\downarrow\rangle_{B}$,
that allows a classical dead \emph{or} alive description. This case
is especially interesting, because it focuses on the steering of a
\emph{mesoscopic} system (the Schrodinger cat) \cite{thermal he andR}.
If we assume Local Realism is valid, then it is the local state of
the mesoscopic system that is shown to be inconsistent with the quantum
mechanics (refer to the LHS model of the Section II.B). This contrasts
with EPR's original argument, which showed the inconsistency for a
\emph{microscopic} system \cite{epr}.

\subsection{Decoherence of the Schrodinger cat with a heat bath}

It will prove useful to next study the interaction of the single mode
``Schrodinger cat'' state \cite{yurke stoler,key-14,key-12,key-13}
\begin{equation}
|\psi\rangle=\frac{1}{\sqrt{2}}\{|-\alpha\rangle+i|\alpha\rangle\}\label{eq:macrosup-2}
\end{equation}
\textcolor{red}{}with a reservoir. This was analysed by many authors,
including Yurke and Stoler \cite{yurke stoler,legdecoh,key-12,key-13,key-14}.
We consider that the single mode, prepared in the Schrodinger cat
state, is then coupled to a thermal heat bath with dissipation. Using
the solutions from Section II.C, we can calculate the moments at a
later time $t$, in terms of the initial moments:\textcolor{red}{}
\begin{eqnarray*}
\left\langle a^{\dagger}(t)a(t)\right\rangle  & = & e^{-2\gamma t}\left\langle a^{\dagger}(0)a(0)\right\rangle +n\left(1-e^{-2\gamma_{a}t}\right)
\end{eqnarray*}
\begin{eqnarray*}
\left\langle a(t)a(t)\right\rangle  & = & e^{-2\gamma t}\left\langle a(0)a(0)\right\rangle 
\end{eqnarray*}
Denoting the variances using shorthand notation, by $\Delta^{2}P=(\Delta P)^{2}$
and $\Delta^{2}X=(\Delta X)^{2}$, we find\textcolor{blue}{}\textcolor{green}{}
\begin{eqnarray}
\Delta^{2}P & = & 1+2n\left(1-e^{-2\gamma t}\right)-4\alpha^{2}e^{-2\gamma t}e^{-4\alpha^{2}}\label{eq:VarP_SecIIB}
\end{eqnarray}
\textcolor{red}{}We see that the variance for $P$ can be reduced
below the quantum limit (given by $1$).\textcolor{black}{{} The reduction
of the variance of $P$ below the quantum limit is in itself a signature
for the ``cat'' state. We see this as follows. If we take $\alpha$
large, and assume no thermal noise and $\gamma=0$, then the probability
distribution $P(x)$ associated with each of the ``dead'' and ``alive''
states is a Gaussian hill, with variance $1$. If the system were
to actually be in any kind of mixture of these quantum states, then
the variance for $P$ could not drop below $1$ because of the uncertainty
relation (and the fact that mixing states cannot decrease the variance)
\cite{erci cav reid pra ,prl eric}.}\textcolor{red}{{} }\textcolor{black}{The
observation of a reduced variance for $P$ can thus occur for a superposition}\textcolor{black}{\emph{
but not for a mixture}}\textcolor{black}{{} of the two quantum states
that possess a distribution $P(x)$ given by the Gaussian hills. The
details are not studied further here however, since our objective
is to study the decoherence of steering.}\textcolor{red}{}\textcolor{blue}{}\textcolor{red}{{}
}
\begin{figure}
\begin{centering}

\par\end{centering}

\begin{centering}
\includegraphics[scale=0.2]{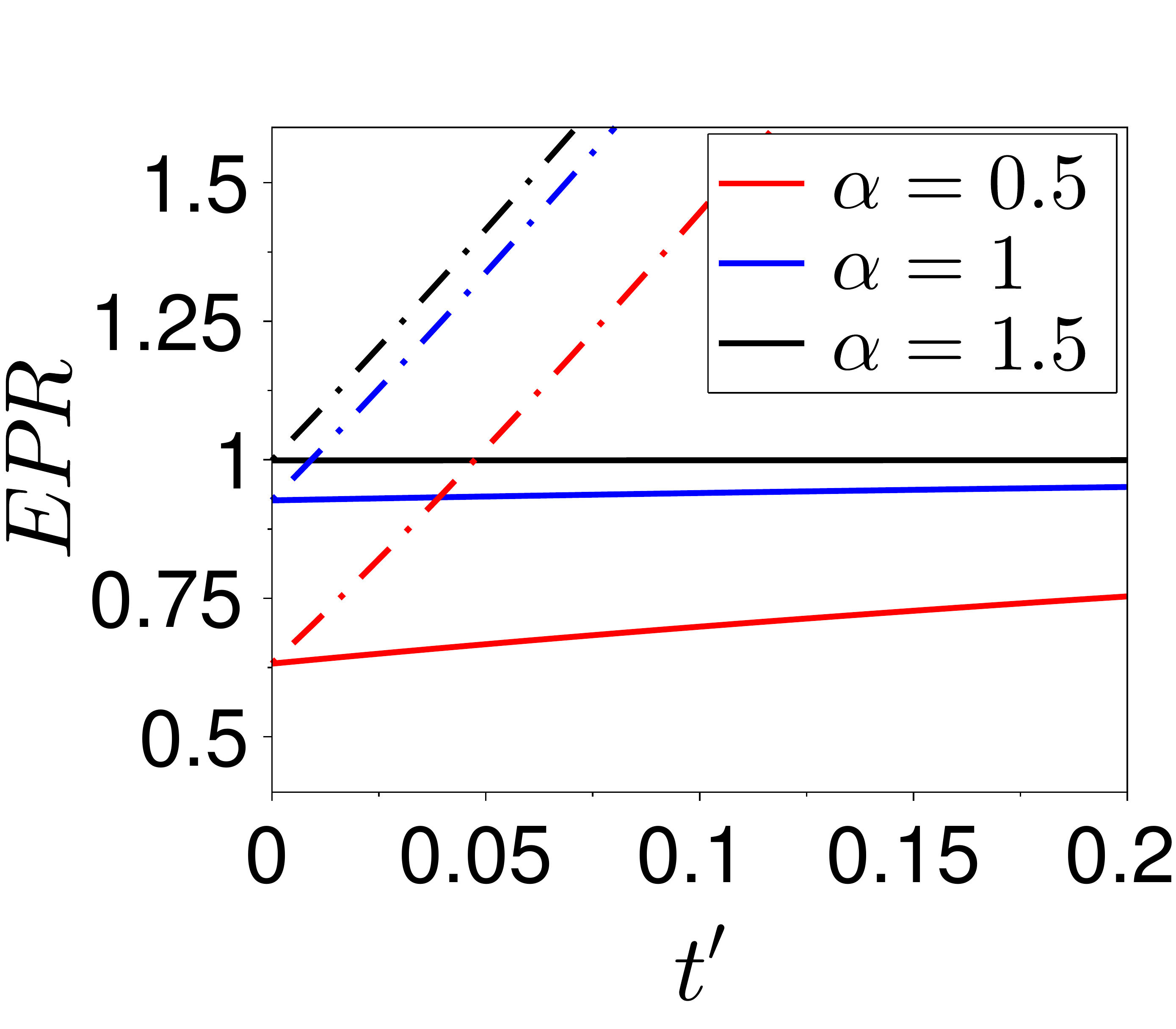}
\par\end{centering}

\protect\caption{\emph{\label{fig:EPR_Decoherence}}The decoherence of the Schrodinger
cat EPR-steering signature $EPR$ for the system depicted in Fig.
7. Here, a thermal reservoir is coupled to the spin system $B$. We
plot the steering signature versus $t'=\gamma t$. Here we consider
different sizes of the ``cat'' (solid line lower to top, $\alpha=0.5,$
$1$ and $1.5$). For each $\alpha$, we plot different values of
$n$: $n=0$ (solid), $n=1$ (dashed-dotted). \textcolor{blue}{}}
\end{figure}

\subsection{Decoherence of the steering of the Schrodinger cat}

Motivated by this, we now examine the decoherence of the EPR steering
of a Schrodinger cat state, as given by the signature Eq. (\ref{eq:s}).
We consider the two-mode case where $A$ is a harmonic oscillator
coupled to a heat bath from time $t=0$ and the second system is the
spin $B$ (no heat bath coupling), as depicted in Figure 7. The system
is \emph{prepared }in the entangled Schrodinger cat state
\begin{equation}
|\psi\rangle=\frac{1}{\sqrt{2}}\{|-\alpha\rangle_{A}|\uparrow\rangle_{B}+i|\alpha\rangle_{A}|\downarrow\rangle_{B}\}\label{eq:macrosup-1-1}
\end{equation}
We can rewrite this state in terms of the basis states $|\uparrow\rangle_{x}$
and $|\downarrow\rangle_{x}$ for the spin $\sigma_{x}$ as shown
in Section III.A. \textcolor{red}{}It is straightforward to calculate
the moments\textcolor{red}{}\textcolor{black}{{} $\langle a\sigma_{z}\rangle$,
$\langle a^{2}\sigma_{z}\rangle$, etc.} at the time $t=0$. The
moments at a later time $t$ (after the interaction with the heat
bath has been turned on) can be evaluated using the solution (\ref{eq:soloplang})
of Section II.C. We note that since the reservoir interaction does
not involve the spins, we have $\sigma_{z}(t)=\sigma_{z}(0)$, $\sigma_{x}(t)=\sigma_{x}(0)$.
Therefore, we can calculate the moments at the time $t$, in terms
of the moments at the initial time $t=0$.

We next consider the moments of the distributions at the time $t$
for $X_{A}$ and $P_{A}$, \emph{conditional} on getting the result
either $+1$ or $-1$ for the Pauli spin measurement on $B$. These
moments can be readily calculated. We outline the method. We use the
notation, for example, $\langle X_{A}\Bigl|1\rangle_{z}$ to denote
the moment $\langle X_{A}\rangle$ of system $A,$ conditioned on
the result $+1$ for spin $\sigma_{z}$ at $B$. Let us suppose then
that we obtain the outcomes $X_{A}$ at $A$, and $\sigma_{z}$ at
$B$. \textcolor{black}{We define the measurable probability $P(X_{A},\sigma_{z})$
for the joint outcomes. We see that: $P(X_{A}\Bigl|\sigma_{z})=\frac{P(X_{A},\sigma_{z})}{P_{z}(\sigma_{z})}$.
It follows that $P(X_{A},\pm1)=P(X_{A}\Bigl|\pm1)P_{z}(\pm1)=\frac{1}{2}P(X_{A}\Bigl|\pm1)$.
Hence, $\langle X_{A}\sigma_{z}\rangle=P(X_{A},1)X_{A}-P(X_{A},-1)X_{A}=P(X_{A}\Bigl|1)P_{z}(1)X_{A}-P(X_{A}\Bigl|-1)P_{z}(-1)X_{A}=\frac{1}{2}\{\langle X_{A}\Bigl|1\rangle-\langle X_{A}\Bigl|-1\rangle\}$.
Following this procedure, we see that:}
\begin{eqnarray}
\langle X_{A}\sigma_{z}\rangle & = & \frac{1}{2}\{\langle X_{A}\Bigl|1\rangle_{z}-\langle X_{A}\Bigl|-1\rangle_{z}\}\nonumber \\
\langle X_{A}\rangle & = & \frac{1}{2}\{\langle X_{A}\Bigl|1\rangle_{z}+\langle X_{A}\Bigl|-1\rangle_{z}\}\label{eq:sp2}
\end{eqnarray}
and similarly for the moments involving $X_{A}^{2}$. This allows
us to solve for the conditional moments using the relations such as
$\langle X_{A}\Bigl|1\rangle_{z}=\langle X_{A}\sigma_{z}\rangle+\langle X_{A}\rangle$
and $\langle X_{A}\Bigl|-1\rangle_{z}=-\langle X_{A}\sigma_{z}\rangle+\langle X_{A}\rangle$.
We also define the same relations, but replacing $X_{A}$ with\textcolor{black}{{}
$P_{A}$ and $\sigma_{z}$ with $\sigma_{x}$. }\textcolor{red}{}The
final solutions for the conditional moments are: $\langle X_{A}\Bigl|\pm1\rangle_{z}=\mp2\alpha e^{-\gamma t}$,
$\langle X_{A}^{2}\Bigl|\pm1\rangle_{z}=1+2n\left(1-e^{-2\gamma t}\right)+4\alpha^{2}e^{-2\gamma t}$,
$\langle P_{A}\Bigl|\pm1\rangle_{x}=\mp2\alpha e^{-\gamma t}e^{-2\alpha^{2}}$
and $\langle P_{A}^{2}\Bigl|\pm1\rangle_{x}=1+2n\left(1-e^{-2\gamma t}\right)$.
Hence we solve for the average variance $Var(X_{A}\Bigl|\sigma_{z})$
of $X$ conditioned on the measurement of spin $\sigma_{z}$ at $B$:
We obtain
\begin{equation}
Var(X_{A}\Bigl|\sigma_{z})=1+2n\left(1-e^{-2\gamma t}\right)\label{eq:xvar}
\end{equation}
 Similarly, we can solve for variance $Var(P_{A}\Bigl|\sigma_{x})$
of $P_{A}$ conditioned on spin $\sigma_{x}$ at $B$, to obtain\textcolor{green}{}
\begin{equation}
Var(P_{A}\Bigl|\sigma_{x})=1+2n\left(1-e^{-2\gamma t}\right)-4\alpha^{2}e^{-2\gamma t}e^{-4\alpha^{2}}\label{eq:pvar}
\end{equation}
 To show how the EPR steering signature decoheres with time, we
evaluate\textcolor{black}{
\begin{equation}
{\color{black}EPR=Var(X_{A}\Bigl|\sigma_{z})Var(P_{A}\Bigl|\sigma_{x})}\label{eq:s-1}
\end{equation}
}Here, we use the notation defined in Section III.A.  

We find that with $n=0$ (no thermal noise), the value of $EPR$ is
identical to that of $Var(P_{A}\Bigl|\sigma_{x}),$ which is also
identical to the variance $\Delta^{2}P$ given by Eq. (\ref{eq:VarP_SecIIB}).
This value is plotted in Figure 8 (solid lines). \textcolor{black}{By
the above argument, the signature of the S-cat is the drop below $1$
of $EPR$. We note that as $\alpha$ increases, the $EPR$ value tends
to $1$, the signature therefore becoming more sensitive to decoherence.
This reduction in variance of $P$ as $\alpha$ increases is directly
associated with the interference fringes in the distribution $P(P_{A})$
\cite{yurke stoler,key-14,key-12,key-13}. These fringes become finer
as $\alpha$ increases, as evident from the function given by Eq.
(\ref{eq:fringe5}). }With thermal noise, we see that the value of
$EPR$ increases: a much greater sensitivity to decoherence is apparent,
as illustrated in Figure 8 by the dashed-dotted curves.\textcolor{blue}{}

\section{Conclusion}

The directional properties of EPR steering are significant when it
comes to understanding the decoherence of EPR-steering. We have shown
how when two systems are coupled independently to a reservoir, the
steering is asymmetrically affected. If we consider the steering of
a system $A$ by measurements made on the system $B$, then the steering
is sensitive to the reservoir coupling to system $B$. This is intuitively
not surprising, given the Local Hidden State (LHS) definition of EPR
steering as a nonlocality. In terms of the LHS description that is
to be negated in order to confirm such steering by B \cite{hw-steering-1,hw2-steering-1},
it is the hidden states of system $B$ that are like those considered
by Bell. The sensitivity of the steering to the losses on the system
$B$ has been studied experimentally, in the context of the EPR paradox
\cite{buono bow,erci cav reid pra ,key-34}. There, it was known that
the EPR criterion could not be achieved, with $50\%$ or more losses
on the steering system $A$. Recent work considers this result in
terms of the monogamy properties of EPR steering \cite{monog}. The
sensitivity of the steering signatures to losses (which may represent
an eavesdropper on the channel) has potentially important implications
for quantum cryptography \cite{onesided,onesidedbog}.

The behaviour with respect to the thermal noise appears almost reversed.
This has implications for detecting steering of Schrodinger cats which
are coupled to hot reservoirs. For the Schrodinger cat example, we
explained how the steering acts as a signature of the Schrodinger
cat. Adding thermal noise to the system $B$ significantly affects
the amount of EPR steering of system $A$. There have been recent
theoretical studies of the steering of a mechanical oscillator, which
we liken to a ``cat'', by measurements made on an optical pulse
\cite{thermal he andR,simon}. In those studies, the thermal noise
of the oscillator was shown to have a quite dramatic effect on the
amount of steering possible. By comparison, the entanglement between
the thermal oscillator and the pulse was quite robust. Similar results
have been obtained for Bose Einstein condensates \cite{karenbec}.
These results are consistent with the simple descriptions of decoherence
given in this paper.
\begin{acknowledgments}
This work was supported by an Australian Research Council Discovery
Project Grant. We are grateful to B. Dalton for many helpful suggestions.\end{acknowledgments}

\end{document}